\documentclass[iop,8pt]{emulateapj}
\usepackage{natbib}
\usepackage[colorlinks,urlcolor=blue,citecolor=blue,linkcolor=blue]{hyperref}
\usepackage[normalem]{ulem}

\newcommand{\rev}[1]{{#1}}
\newcommand{\revv}[1]{{#1}}
\newcommand{\revvv}[1]{{#1}}
\newcommand{\AB}{}

\newcommand{\eqb}{\begin{eqnarray}}
\newcommand{\eqe}{\end{eqnarray}}
\newcommand{\be}{\begin{eqnarray}}
\newcommand{\ee}{\end{eqnarray}}
\newcommand{\bi}{\begin{itemize}}
\newcommand{\ei}{\end{itemize}}
 \newcommand{\gammab}{\gamma}
\newcommand{\betab}{\beta}
\newcommand{\betabx}{\beta_x}

\newcommand{\upl}{u}

\newcommand{\gammacr}{\gamma_{\rm cr}}
\newcommand{\etarec}{\eta_{\rm rec}}

\newcommand{\unit}[1]{\nobreak{\mathrm{\;#1}}} % For units of measure within math mode

\def\comp{\,c/\omega_{\rm p}}

\newcommand{\sect}[1]{Sect.~\ref{sec:#1}}

\newcommand{\fig}[1]{Fig.~\ref{fig:#1}}
\newcommand{\eq}[1]{Eq.~(\ref{eq:#1})}

\newcommand{\rhot}{\,r_{0,\rm hot}}
\newcommand{\alf}{Alfv\'en}

\newcommand{\gh}{{\gamma_e^\prime}}
\newcommand{\game}{\gamma_e}
\newcommand{\Tb}{T_{\rm b}}
 \newcommand{\fHE}{f_{\rm HE}}

\usepackage{graphicx}
\usepackage{epsfig} 
\usepackage{epstopdf}

\newcommand{\bmath}[1]{\mbox{\boldmath{$#1$}}}

\begin{document}
\newcommand{\note}[1]{{\bf \color{blue} [ MP #1]}}
\title[]
{Kinetic Simulations of Radiative 
Magnetic
Reconnection in the 
Coronae of Accreting Black Holes}

\author{Lorenzo Sironi$^1$\thanks{E-mail: lsironi@astro.columbia.edu.} 
and Andrei M. Beloborodov$^{2,3}$}
\affil{
$^1$Department of Astronomy and Columbia Astrophysics Laboratory, Columbia University, 550 W 120th St, New York, NY 10027, USA\\
$^2$Physics Department and Columbia Astrophysics Laboratory, Columbia University, 550 W 120th St, New York, NY 10027, USA \\
$^3$Max Planck Institute for Astrophysics, Karl-Schwarzschild-Str. 1, D-85741, Garching, Germany}

\begin{abstract}
We perform two- and \rev{three-dimensional} particle-in-cell simulations of reconnection in magnetically dominated $e^\pm$ plasmas subject to strong Compton cooling.
Magnetic reconnection under such conditions can operate in accretion disk coronae around black holes, which produce hard X-rays through Comptonization. Our
simulations 
show 
that most of the plasma in the reconnection layer is 
kept cold by Compton losses and locked in magnetically dominated plasmoids with a small thermal pressure.
Compton drag clears cavities inside plasmoids and also affects their bulk motions. These effects, however, weakly change the reconnection rate and the plasmoid size distribution 
from those in non-radiative reconnection. 
This demonstrates that the reconnection dynamics is governed by similar magnetic stresses in both cases and weakly affected by thermal pressure.
We 
examine the energy distribution of particles energized by radiative reconnection 
and observe
two distinct components. 
(1) A mildly relativistic peak, which  results from 
bulk motions 
of cooled
plasmoids. This component receives most of the dissipated reconnection power and dominates the output X-ray 
emission.
The peak has a quasi-Maxwellian shape with an effective temperature of $\sim 100$~keV. Thus, it mimics thermal Comptonization used previously to fit hard-state spectra of accreting black holes.
(2) A high-energy tail, which 
receives $\sim 20$\% of the dissipated reconnection power.
It is populated by particles accelerated impulsively at X-points or ``picked up'' by 
fast outflows from X-points.
The 
high-energy
particles immediately cool, and their inverse Compton emission explains the MeV spectral tail detected in the hard state of Cyg X-1. Our first-principle simulations support magnetic reconnection as a mechanism powering hard X-ray emission from \revv{magnetically dominated regions} of accreting black holes.  
\end{abstract} 

\maketitle

\section{Introduction}
Luminous accretion disks around stellar-mass black holes are routinely observed in ``soft'' and ``hard'' X-ray states \citep[e.g.,][]{zdz_04}. The soft state is dominated by quasi-thermal emission from an optically thick accretion disk, while the hard state is usually explained by a phenomenological
model of Comptonization of soft X-rays
in a hot plasma (``corona'') with electron temperature of $\sim100$ keV. 
The electron energization mechanism required to balance the fast inverse Compton (IC) losses of the corona is unknown. Recently, \citet{belo_17} (hereafter B17) proposed that 
 this mechanism is the bulk acceleration of multi-scale plasmoids generated by magnetic reconnection.

  The corona above the accretion disk has a low mass density $\rho$ and is likely dominated by the disk magnetic field $B$, with the magnetization parameter $\sigma\equiv B^2/4\pi\rho c^2>1$. Then the \alf\, speed in the corona is close to the speed of light and magnetic reconnection proceeds
in the relativistic regime. 
  This regime was explored analytically \citep[e.g.,][]{lyutikov_uzdensky_03,lyubarsky_05} and 
studied with
fully kinetic particle-in-cell (PIC) simulations. The simulations provided a detailed view 
 of relativistic reconnection, 
   including reliable measurements of 
the reconnection 
  rate, and revealed
the mechanisms of particle acceleration (see \citealt{kagan_15} for a review). With unprecedentedly large-scale simulations, \citet{sironi_16} have investigated the properties of the chain of plasmoids (magnetic islands) that are copiously generated in the reconnection layer, as a self-consistent by-product of the system evolution 
\citep{uzdensky_10}. 

For conditions appropriate to coronae of accreting black holes, 
B17 argued that the plasmoid chain experiences fast IC losses accompanied by electron-positron pair creation,
   estimated the resulting distribution of plasmoid bulk speeds, and performed Monte-Carlo simulations of photon Comptonization by the plasmoid motions. 
The self-regulated plasmoid chain was found
to radiate its energy in hard X-rays with a spectrum similar to the hard-state spectra observed in black hole X-ray binaries and AGN.

  In the present work, we employ kinetic PIC simulations to study  relativistic magnetic reconnection with strong IC losses. Numerous previous 
studies of relativistic electron-positron reconnection with
the PIC technique focused on the regime with negligible cooling 
\citep[e.g.,][]{zenitani_01,lyubarsky_liverts_08,kagan_13,ss_14,guo_14,guo_15a,guo_19,nalewajko_15,werner_16,werner_17,sironi_15,sironi_16,petropoulou_18}.
One exception is the models of reconnection in pulsars and pulsar wind nebulae, where synchrotron emission from extremely energetic electrons was included in the PIC simulations 
\citep[e.g.,][]{cerutti_13a,cerutti_13b,kagan_16,kagan_18,hakobyan_18}. 
Recently, 
\citet{werner_19} incorporated IC effects in their PIC simulations
(see also \citealt{nalewajko_18}), and showed that radiative cooling considerably reduces the efficiency of relativistic reconnection in producing nonthermal particles. 

In this work, we present a systematic investigation of the role of strong IC losses on the plasmoid chain dynamics and 
nonthermal particle acceleration. 
We perform large-scale two- and \rev{three-dimensional} PIC simulations of relativistic reconnection in $e^\pm$ plasma immersed in a dense radiation field, which exerts Compton drag on the particles. 
   For simplicity, we assume that the radiation field is fixed during the simulation and composed of low-energy photons, so that Compton scattering occurs in the Thomson regime. The efficiency of Compton drag is determined by the radiation density, which is the main parameter of the problem, besides the magnetization parameter $\sigma$.
We employ outflow boundary conditions along the reconnection exhausts;
then the plasmoid chain dynamics can be reliably studied in a quasi-steady state.  A small guide field is included in the setup of the reconnection layer. 
     
This paper is organized as follows. In \sect{setup} we describe the setup of our simulations and our choice for parameterization of the radiation 
density. In \sect{params} we justify the physical parameters of our simulations as appropriate for the coronae of accreting black holes. In \sect{results} we present our 2D results, focusing on the dynamics of the plasmoid chain and the spectrum of post-reconnection particles. 
   We compare a representative run with strong IC losses with a no-cooling simulation (which otherwise has identical parameters), and then show how various properties of the reconnection system depend on the degree of IC cooling by varying it from zero to extreme values. \rev{In \sect{3d} we show that our main findings also hold in 3D simulations.} Our conclusions and observational implications are summarized in \sect{disc}.

%%%%%%%%%%%%%%%%%%
\section{Simulation setup}\label{sec:setup}
We use the 3D electromagnetic PIC code TRISTAN-MP \citep{buneman_93, spitkovsky_05} to study relativistic reconnection in electron-positron (pair) plasmas. 
% Our reference 
\AB{Most of our}
simulations employ a 2D spatial domain in the $xy$ plane 
\AB{except the full 3D simulations presented in \sect{3d}. In all simulations}  
 we track all three components of 
\AB{the particle velocities}
and of the electromagnetic fields.  Our 2D simulation setup closely parallels 
\AB{that}
in \citet{sironi_16}.
\AB{It is summarized below}
 for completeness.  
 \AB{Details of the 3D setup are deferred to \sect{3d}.}

The reconnection layer is set up in Harris equilibrium, with the initial magnetic field $\bmath{B}_{\rm in}=-B_0\, \bmath{\hat{x}}\tanh\,(2\pi y/\Delta)$  reversing at $y=0$ over a thickness $\Delta$ that will be specified below. The field strength is parameterized by the magnetization 
\AB{
\begin{equation}
  \sigma=\frac{B_0^2}{4\pi m_e n_0 c^2}=\frac{\omega_{\rm c}^2}{\omega_{\rm p}^2}, 
\end{equation}
}
where $\omega_{\rm c}=eB_0/m_ec$ is the Larmor frequency and $\omega_{\rm p}=\sqrt{4\pi n_0 e^2/m_e}$ is the plasma frequency for the cold electron-positron plasma outside the layer ($n_0$ is the number density, including both species), which is initialized with a small thermal spread of $kT/m_e c^2=10^{-4}$.
The \alf\ speed is related to the magnetization as $v_A/c=\sqrt{\sigma/(\sigma+1)}$.
We focus on the regime $\sigma\gg1$ (i.e., $v_A/c\sim 1$) of relativistic reconnection, and take $\sigma=10$ as our fiducial case.  In addition to the reversing field, we initialize a uniform component of ``guide field'' along $z$ (i.e., aligned with the electric current in the reconnection layer) with strength $B_g=0.1\,B_0$. 
The guide field 
provides pressure support to the cores of strongly-cooled 
(and somewhat compressed)
 plasmoids. An investigation of the dependence of our results on $\sigma$ and $B_g/B_0$ is left for a future work.

Magnetic pressure outside the current sheet is 
initially
balanced by particle pressure in the sheet, by adding a component of hot plasma that is overdense 
by a factor of 3
relative to the number density $n_0$ of cold particles outside the layer. All the quantities presented in \sect{results} and \sect{3d} (e.g., thermodynamic properties of the layer and particle energy spectra) do not include 
this particle population set up in the current sheet, since its properties depend on arbitrary choices at initialization.
The hot component is used only to setup the initial reconnection. After approximately one light-crossing time of the box, the artificially hot particles get ejected from the system. Then reconnection proceeds in a quasi-steady state, losing memory of the initial hydrostatic state with the artificial hot component.

We trigger reconnection near the center of the computational domain, by removing the pressure of the hot particles initialized in the current sheet. This triggers a local collapse of the layer, which generates an X-point at the center.
The initial perturbation results in the  formation of two ``reconnection fronts'' that propagate away from the center along $\pm\bmath{\hat{x}}$ (i.e., along  the current layer), at roughly the \alf\  speed. We choose the thickness of the current sheet $\Delta$ to be large enough such that reconnection does not get spontaneously triggered anywhere else in the current layer, outside of the region in between the two reconnection fronts.\footnote{For the same reason, we prescribe by hand that the particles belonging to the hot population initialized in the layer are not subject to IC losses.} 
In units of 
$\rhot=\sqrt{\sigma}\comp$ 
(which corresponds to the Larmor radius of particles with energy $\sigma m_e c^2$ in 
field $B_0$)\footnote{If reconnection were to transfer all of the field energy to the particles, the mean particle energy would be $\sim \sigma m_e c^2/2$. So, our definition of $\rhot$ corresponds, apart from a factor of two, to the Larmor radius of the typical particles heated/accelerated by reconnection.}
the thickness $\Delta$ is chosen to be $\Delta\sim30\rhot\sim 100\comp$.

After one \alf ic crossing time ($=L/v_A$, where $L$ is the half-length of our box along the $x$ direction), the two reconnection fronts reach the $x$ boundaries of the computational domain. Here, we employ absorbing boundary conditions in the $x$ direction of the reconnection outflow, to mimic an open boundary where no information is able to propagate back inward \citep{daughton_06,cerutti_15,belyaev_15}. We refer to \citet{sironi_16} for a discussion of our implementation of outflow boundaries. Along the $y$ direction of the reconnection inflow, we employ two ``moving injectors'' (receding from $y=0$ along $\pm {\bmath{\hat{y}}}$) and an expanding simulation box, a technique that we have extensively employed in our studies of relativistic shocks \citep{sironi_spitkovsky_09,sironi_spitkovsky_11a,sironi_13} and relativistic reconnection \citep[e.g.,][]{ss_14,sironi_16,petropoulou_18}.
The two injectors constantly introduce fresh magnetized  plasma into the simulation domain. This permits us to evolve the system as far as the computational resources allow, retaining all the regions that are in causal contact with the initial setup. 

In 
\AB{the}
2D simulations, we typically employ four particles per cell (including both species). 
\AB{We have checked that the results are weakly changed}
when using 16 particles per cell (extensive tests in the uncooled case have also been performed in \citealt{sironi_16}). In order to reduce noise in the simulations, we filter
the electric current deposited by the particles onto the grid, effectively mimicking the role of a larger number of particles per cell \citep{spitkovsky_05,belyaev_15}.
\AB{The plasma skin depth $\comp$ is resolved with 5 cells,}
 so that the Larmor gyration period $2\pi/\omega_{\rm c}=2\pi/\sqrt{\sigma}\,\omega_{\rm p}$ is resolved with at least a few timesteps (the numerical speed of light is 0.45 cells/timestep).

In order to properly extrapolate our results to astrophysical  scales, large-scale computational domains are essential. In the following, we shall call $L$ the half-length of the computational domain along the $x$ direction, i.e., along the reconnection layer. In units of the Larmor radius of hot particles $\rhot=\sqrt{\sigma}\comp$, the half-length $L$ of our fiducial 2D runs is $L\simeq 1062\rhot$. This corresponds to $L/(\comp)\simeq3360$, or equivalently 33,600 cells for the  overall box size along the $x$ direction (so, for the full length $2\,L$).\footnote{The box size along $y$ increases over time, and at the end it is comparable or larger than the $x$ extent.} We evolve our 
simulations up to $\sim 5\,L/c$, corresponding to $\sim 185,000$ timesteps. This 
\AB{provides sufficient statistics to study}
the steady state of the system, which is established after $\sim L/c$, when the reconnection fronts reach the $x$ boundaries.
For the two 
fiducial cases discussed in \sect{results}, one without IC cooling losses and the other with strong cooling, we have also performed larger simulations, with $L\simeq 2125\rhot\simeq 6720\comp$ (as well as a number of smaller simulations, which we do not report in this paper).

\subsection{Compton drag}\label{sec:cd}

In addition to the standard Lorentz force, the particles in our simulations are subject to 
   ``Compton drag'' force,
\be
\label{eq:F}
\bmath{F}_{\rm IC} = -\frac{4}{3}\sigma_{\rm T} \gamma_e^2 U_{\rm rad}\,\bmath{\beta}_e,
\ee
where $\bmath\beta_e$ is the particle velocity in units of the speed of light $c$, 
   $\gamma_e=(1-\beta_e^2)^{-1/2}$ is its 
Lorentz factor, $\sigma_{\rm T}$ 
   is 
the Thomson cross section and $U_{\rm rad}$ 
   is
the 
   radiation
energy density. 
   Equation~(\ref{eq:F}) assumes that the radiation is 
isotropic in the simulation frame (the frame of the reconnection layer, in which the net Poynting flux  vanishes).
   It also assumes that $U_{\rm rad}$ is made of a large number of low-energy photons that scatter in the Thompson regime. This implies that the particle 
loses 
a small fraction of its energy 
   in each scattering event, and the dynamic effect of scattering may be approximated as a continuous drag or friction opposing the particle motion. 

   One can define a characteristic Lorentz factor $\gammacr\gg 1$ at which the maximum electric field in the reconnection layer, $E_{\rm rec}=\eta_{\rm rec}B_0$ ($\eta_{\rm rec}\sim 0.1$), is just sufficient to balance the Compton drag,
\be\label{eq:gammacr1}
    e E_{\rm rec}=\frac{4}{3}\sigma_{\rm T} \gammacr^2 U_{\rm rad}~.
\ee
   Particle acceleration beyond $\gammacr$ is suppressed by prohibitive IC losses. 
   Instead of $U_{\rm rad}$, it is convenient to use $\gammacr$ as the parameter controlling Compton drag; $\gammacr=\infty$ corresponds to $U_{\rm rad}=0$ while a low $\gammacr$ corresponds to a large $U_{\rm rad}$ and a strong drag. 
In this work, we 
   run models with $\gammacr$ as low as 11.3. This lowest $\gammacr$ is comparable to $\sigma=10$ chosen in our simulations, which represents a characteristic Lorentz factor for particle acceleration at X-points. Our fiducial model has $\gammacr=16$. For comparison, we also run a benchmark model with $\gammacr=\infty$ (no Compton drag).

   A particle with a Lorentz factor $\gamma_e$ is decelerated by Compton drag on the timescale $t_{\rm IC}=\gamma_e\beta_e m_ec/F_{\rm IC}$, which may be rewritten as
\be
\label{eq:tIC}
  t_{\rm IC}(\gamma_e)=\frac{3m_ec}{4\sigma_{\rm T} U_{\rm rad}\gamma_e}= \frac{1}{\omega_c}\,\frac{\gammacr^2}{\eta_{\rm rec}\gamma_e}.
\ee
Our temporal resolution $\Delta t=0.09\,\omega_{\rm p}^{-1}=0.09\,\sigma^{1/2} \omega_c^{-1}$
    safely resolves this timescale even at the highest $\gamma_e\sim \gammacr$, since
\be
   \frac{\Delta t}{t_{\rm IC}(\gammacr)}\sim 10^{-2}\,\frac{\sigma^{1/2}}{\gammacr}\ll 1.
\ee
This constraint is easily satisfied 
   as long as
$\gammacr\gtrsim \sigma\gg1$, i.e., in the parameter regime we investigate here.

%%%%%%%%%%%%%%%%%%%%

\section{Plasma conditions in black hole coronae}\label{sec:params}

In this section, we discuss the plasma conditions expected in coronae of black hole accretion disks (we refer to B17 for further details). Magnetic loops above the disk are sheared by differential rotation, generating current sheets. We focus on a current sheet of characteristic length $L\sim r_g$, where $r_g=2 G M_{\rm BH}/c^2$ is the Schwarzschild radius of the black hole. The typical thickness of the current sheet is $h\sim \eta_{\rm rec} \,L\sim 0.1 \,L$, which corresponds to the width of the largest plasmoids in the reconnection layer \citep{sironi_16}.

   The reconnection process is initiated in the current sheet and dissipates magnetic energy, similar to solar flares but with a much higher power. This energy release is accompanied by creation of $e^\pm$ pairs inside and around the reconnection layer (see Figure~1 in B17). The pairs can strongly dominate the plasma density, depending on the compactness parameter of the magnetic flare. The pair plasma 
is kept at the Compton temperature of the radiation field
until it flows into the reconnection layer.
For 
simplicity in this work we neglect the ion component and assume a pure electron-positron composition with a given (pre-reconnection) density $n_0$. We also idealize the problem by assuming a zero Compton temperature of the radiation field, so the plasma is cold before being heated by reconnection.

In the radiative regime, most of the dissipated magnetic energy is quickly converted to radiation, which escapes at speed $\sim c$, as long as the Thomson optical depth of the reconnection layer 
$\tau_{\rm T}=n_0 \sigma_{\rm T} h\sim n_0 \sigma_{\rm T}\eta_{\rm rec}L$
is not much greater than unity ($\tau_{\rm T}\sim 1$ is typically inferred for black hole coronae, and such optical depths may be sustained by pair creation in magnetic flares, see B17). 
Thus, energy conservation
implies that the radiation 
density is $U_{\rm rad}\sim \eta_{\rm rec} U_B\sim 0.1\,U_B$,
   where $U_B=B_0^2/8\pi$ is the 
   energy density of the reconnecting magnetic field (i.e., excluding the guide field, which suffers no dissipation).

The magnetization 
   parameter $\sigma$ can be expressed as
\be\label{eq:sigma}
\sigma=\frac{2 U_B}{n_0 m_e c^2}\sim \frac{2\eta_{\rm rec}\ell_B}{\tau_{\rm T}}
\ee
   where 
$\ell_B$ is the magnetic compactness, defined as 
\be
\ell_B=\frac{U_B \sigma_{\rm T} L}{m_e c^2}
\ee
in analogy with the 
   radiation
compactness $\ell_{\rm rad}=U_{\rm rad} \sigma_{\rm T} L/m_e c^2$.

\revvv{Note that in the presence of ions, $\sigma$ can be significantly reduced, as \eq{sigma} would include the ion mass density. Accretion disk coronae can have ion-dominated regions with $\sigma<1$ and pair-dominated regions with $\sigma\gg 1$. GRMHD simulations \citep[e.g.,][]{jiang_19,chatterjee_19} demonstrate the existence of coronal regions (where magnetic pressure dominates the gas pressure) with $\sigma<1$. Regions with $\sigma\gg 1$ likely exist near the black hole spin axis, although they are harder to reproduce in MHD simulations, which employ density floors for numerical stability. Magnetic reconnection can generate heat in both regions, $\sigma<1$ and $\sigma>1$, with a rate scaling as $(1+1/\sigma)^{-1/2}$. It is not known which region dominates the observed hard X-ray emission. In this paper, we focus on the pair-dominated plasma with $\sigma\gg 1$; radiative reconnection in low-magnetization plasmas with ions will be studied elsewhere.}

   A characteristic compactness for black holes accreting at a significant fraction of the Eddington limit is $\ell_B\sim m_p/m_e$ (B17). Then using 
$\eta_{\rm rec}\sim 0.1$ and $\tau_{\rm T}\sim 1$, \eq{sigma} gives $\sigma\sim 400$. {Then reconnection occurs in the ultra-relativistic regime, and the fastest plasmoids are pushed to Lorentz factors} $\sim \sqrt{\sigma}\gg1$, i.e., reconnection 
can give 
ultra-relativistic bulk motions.
    The characteristic energy for particle acceleration at X-points is comparable to the mean energy per particle released in reconnection, $\gamma_X\sim \sigma/4$ (the magnetic energy per particle is $(\sigma/2) m_ec^2$, and another factor of 1/2 takes into account that only about half of the magnetic energy is dissipated, see \citealt{sironi_15}).

Given that $U_{\rm rad}\sim \eta_{\rm rec} U_B$, we can quantify the expected value of $\gammacr$. 
    Using the relation $\sigma_{\rm T}=8 \pi r_e^2/3 $, where $r_e=e^2/m_e c^2$ is the classical electron radius, we find from \eq{gammacr1}
\be
\gammacr=\left(\frac{27 \,L}{16 \ell_B r_e}\right)^{1/4}\sim 10^4 \left(\frac{M_{\rm BH}}{10M_\odot}\right)^{1/4}
\ee
where we have 
   substituted 
$\ell_B\sim m_p/m_e$. 
   The high value of $\gammacr\gg \gamma_X\sim \sigma/4$ implies that particle acceleration 
at X-points 
   is not impeded by Compton drag. The same condition can be expressed by comparing the IC timescale (\eq{tIC}) with the timescale 
  $t_X$ for particle acceleration at X-points,
\be
  \frac{t_{\rm IC}(\gamma_X)}{t_X}=\frac{\gammacr^2}{\gamma_X^2}\gg 1,
  \qquad t_X =\frac{\gamma_X m_e c}{e E_{\rm rec}}.
\ee

On the other hand, the IC cooling timescale is much shorter than the timescale  for plasma advection along the layer $t_{\rm adv}\sim L/v_A\sim L/c$. Their ratio is
\be
\frac{t_{IC}(\gamma_e)}{t_{\rm adv}}=\frac{3}{4 \gamma_e \ell_{\rm rad}} \ll 1.
\ee
This condition must be satisfied in our simulations for any $\game$, including $\game\sim 1$, which requires a large size $L$ of the computational box,
\be\label{eq:lbox}
\frac{L}{\comp}\gg \frac{\gammacr^2}{\eta_{\rm rec} \sqrt{\sigma}}~~.
\ee

In summary, in order to mirror the conditions expected in black hole coronae, our simulations need to have ({\it i}) $\sigma\gg1$, and our choice of $\sigma=10$ is then suitable; ({\it ii}) $\gammacr\gg \gamma_X\sim \sigma/4$, which is satisfied by all our simulations, and
({\it iii}) $L/(\comp)\gg \gammacr^2/(\eta_{\rm rec} \sqrt{\sigma})$. 
This is achieved e.g. in our fiducial 2D model by choosing $\gammacr=16$ and $L/(\comp)\simeq 3360$. 
Thus, although our runs do not have the 
huge $\gammacr\sim 10^4$ expected in black hole coronae, they satisfy the required hierarchy of relevant scales and energies.

%%%%%%%%%%%%%%%%%%%%%%%%%%%%%%%%

\begin{figure*}
\centering
\resizebox{\hsize}{!}{\includegraphics{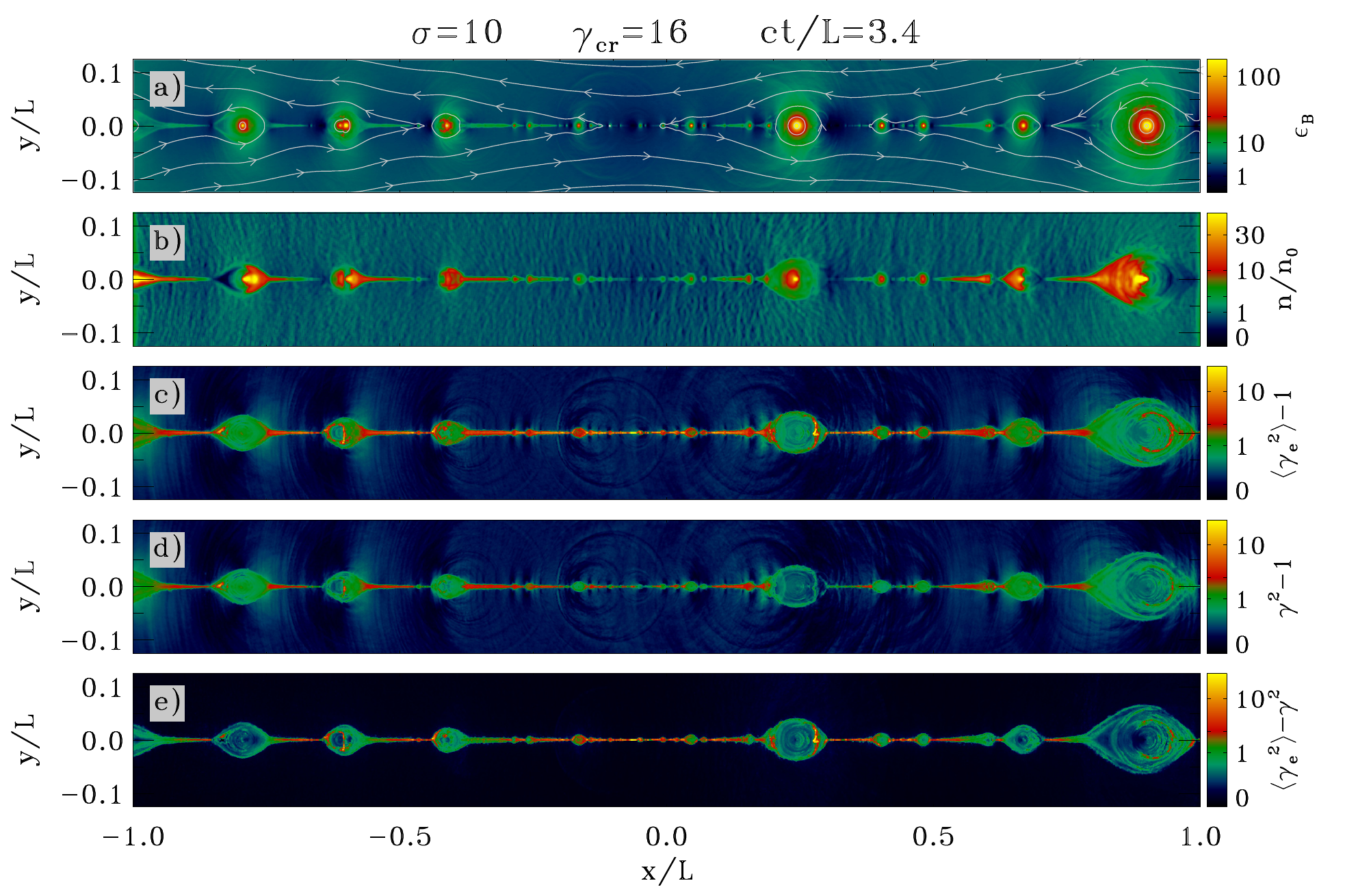}}
 \caption{
 2D structure of the reconnection layer at time $t=3.4\,(L/c)$ in our fiducial run with strong IC losses ($\gamma_{\rm cr}=16$).
We show the region $|y|/L<0.125$ where reconnection occurs 
(the actual extent of the computational box along $y$ grows with time as described in \sect{setup}). 
(a) Magnetic energy density $B^2/8\pi$ normalized by the initial plasma rest-mass energy, $\epsilon_{\rm B}=B^2/8\pi n_0 m_e c^2$, with overplotted magnetic field lines. The initial $\epsilon_B$ far from the reconnection layer is $\epsilon_B\approx \sigma/2=5$. (b) The particle number density $n$, in units of the number density $n_0$ far from the reconnection layer. (c) Local average $\langle\gamma_e^2\beta_e^2\rangle=\langle\gamma_e^2\rangle-1$, obtained for each cell by averaging over the particles in the neighboring $5\times5$ cells.
This quantity is proportional to the local {\it total} IC power radiated per particle. (d) $\gammab^2-1$, where $\gammab$ is the plasma bulk Lorentz factor, as defined in the text. This quantity is proportional to the local {\it bulk} IC power per particle. (e)  The difference $\langle\gamma_e^2\rangle-\gammab^2$ 
represents the contribution from internal particle motions. One can see that the IC power radiated by isolated plasmoids is dominated by their bulk motion 
whereas internal particle motions dominate in the thin regions near the midplane X-points and in the transient reconnection layers formed between merging plasmoids.
}
 \label{fig:fluid2}
\end{figure*}

%%%%%%%%%%%%%%%%%%%

\begin{figure*}
\centering
\resizebox{\hsize}{!}{\includegraphics{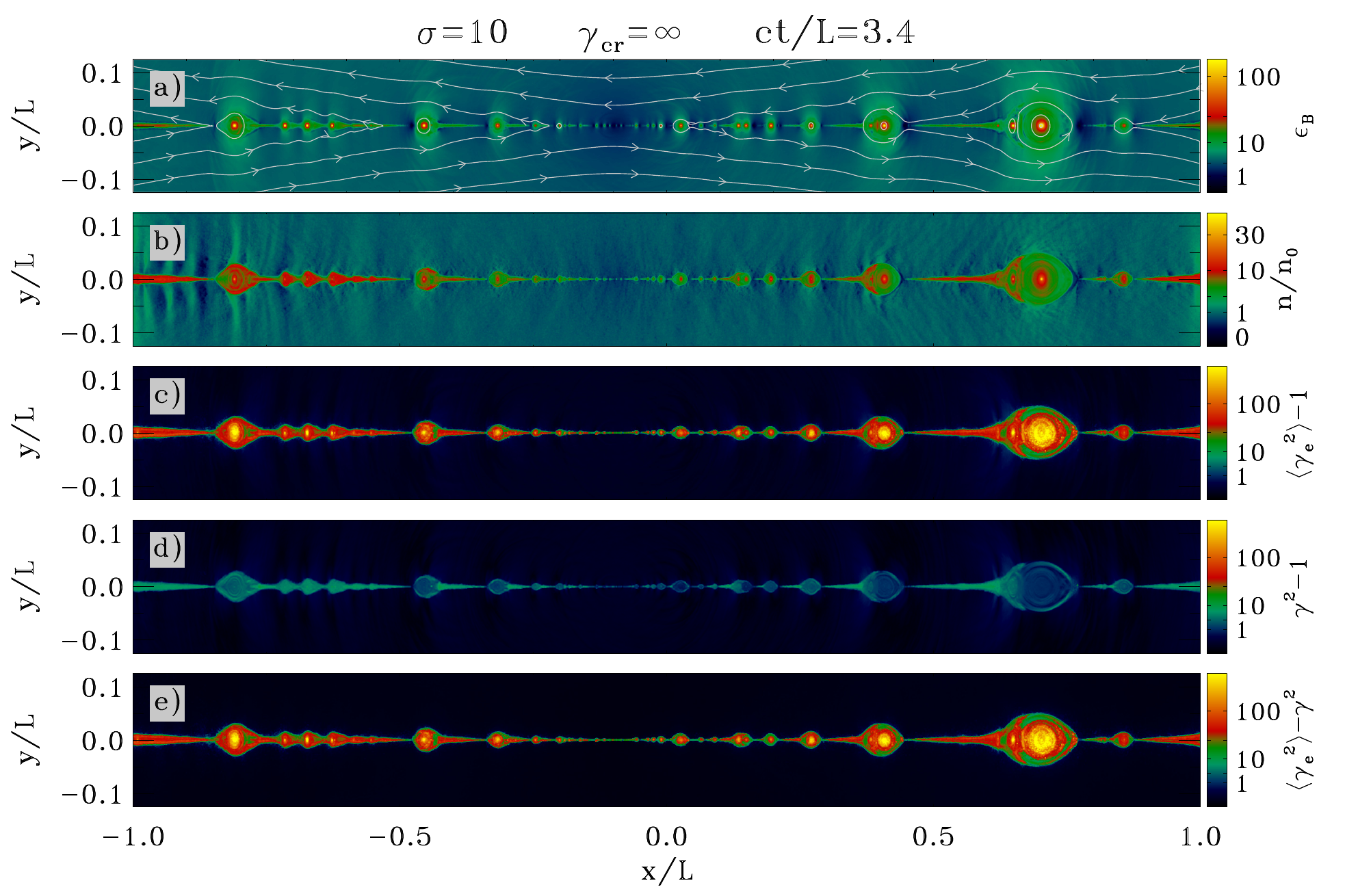}}
 \caption{Same as \fig{fluid2} but for a simulation with no IC losses ($\gammacr=\infty$). Apart from $\gammacr$, all parameters of the simulation are identical to that in \fig{fluid2}, and the snapshot is taken at the same time $t=3.4\,(L/c)$. Internal particle motions now dominate over bulk motions, because particles keep the heat received in reconnection rather than radiate it. However, the overall shape of the plasmoid chain and the reconnection rate are remarkably similar to the run with strong IC losses.}
 \label{fig:fluid1}
\end{figure*}

%%%%%%%%%%%%%%%%%%%%%
\section{Results}\label{sec:results}
In this section, we present 
\AB{the results of our 2D simulations.}
In sections \sect{flow}--\sect{spect}, we compare 
two 
simulations:
one with no IC losses ($\gammacr=\infty$) and the other with strong losses ($\gammacr=16$). We first present the overall flow structure in \sect{flow}, then discuss the effect of IC drag on the plasmoid chain dynamics in \sect{plasm}, and finally explore the particle energy spectrum in  \sect{spect}. 
\sect{scan} presents the results of a suite of 2D simulations aimed at demonstrating
how 
basic parameters of the reconnection layer
depend on $\gammacr$ for a wide range
between
$\gammacr=11.3\sim \sigma$ and $\gammacr=\infty$.

%%%%%%%%%%%%%%%%%%%%%
\begin{figure*}
\centering
{\includegraphics[width=0.45\textwidth]{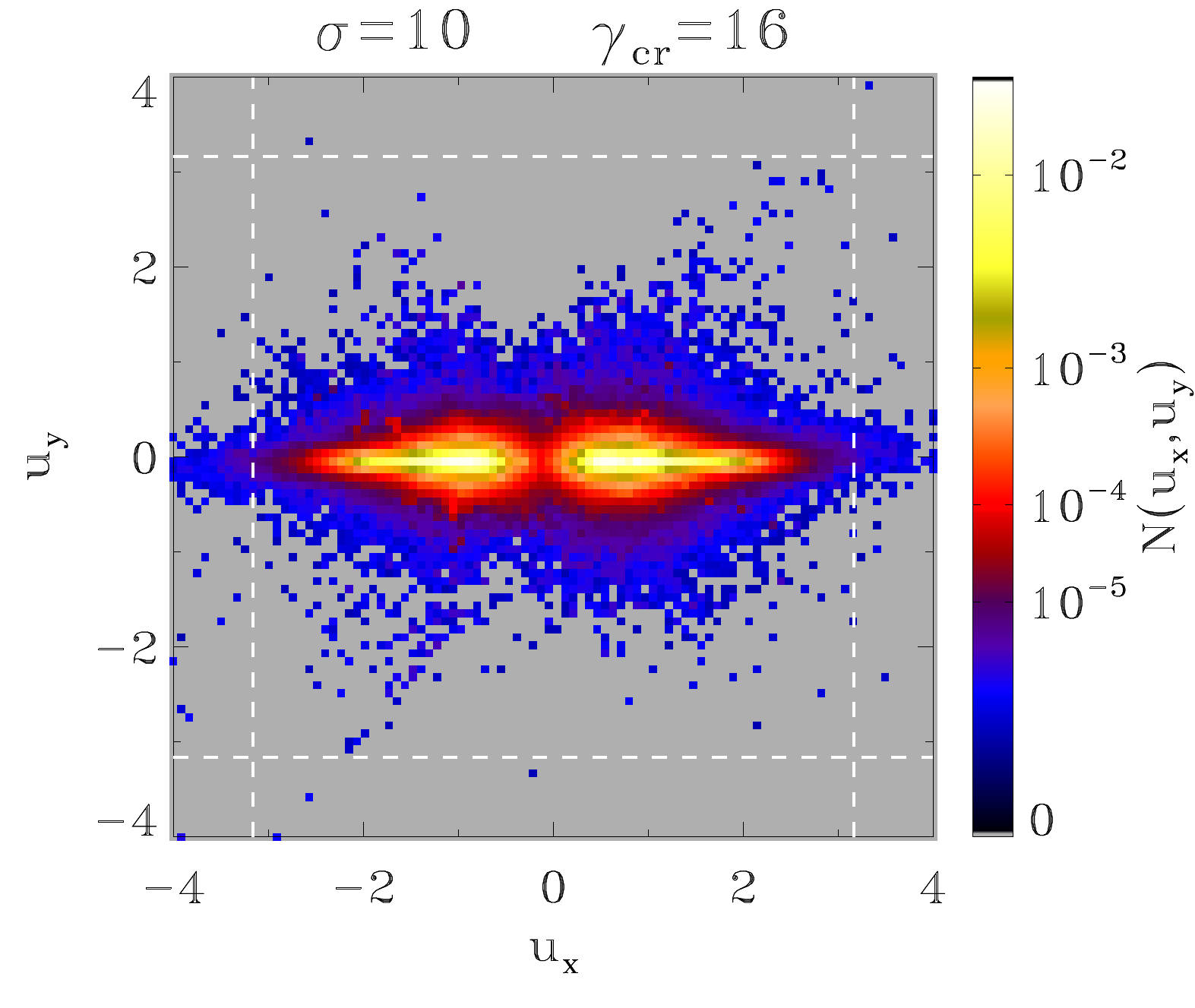}}
\hspace{0.3in}
{\includegraphics[width=0.45\textwidth]{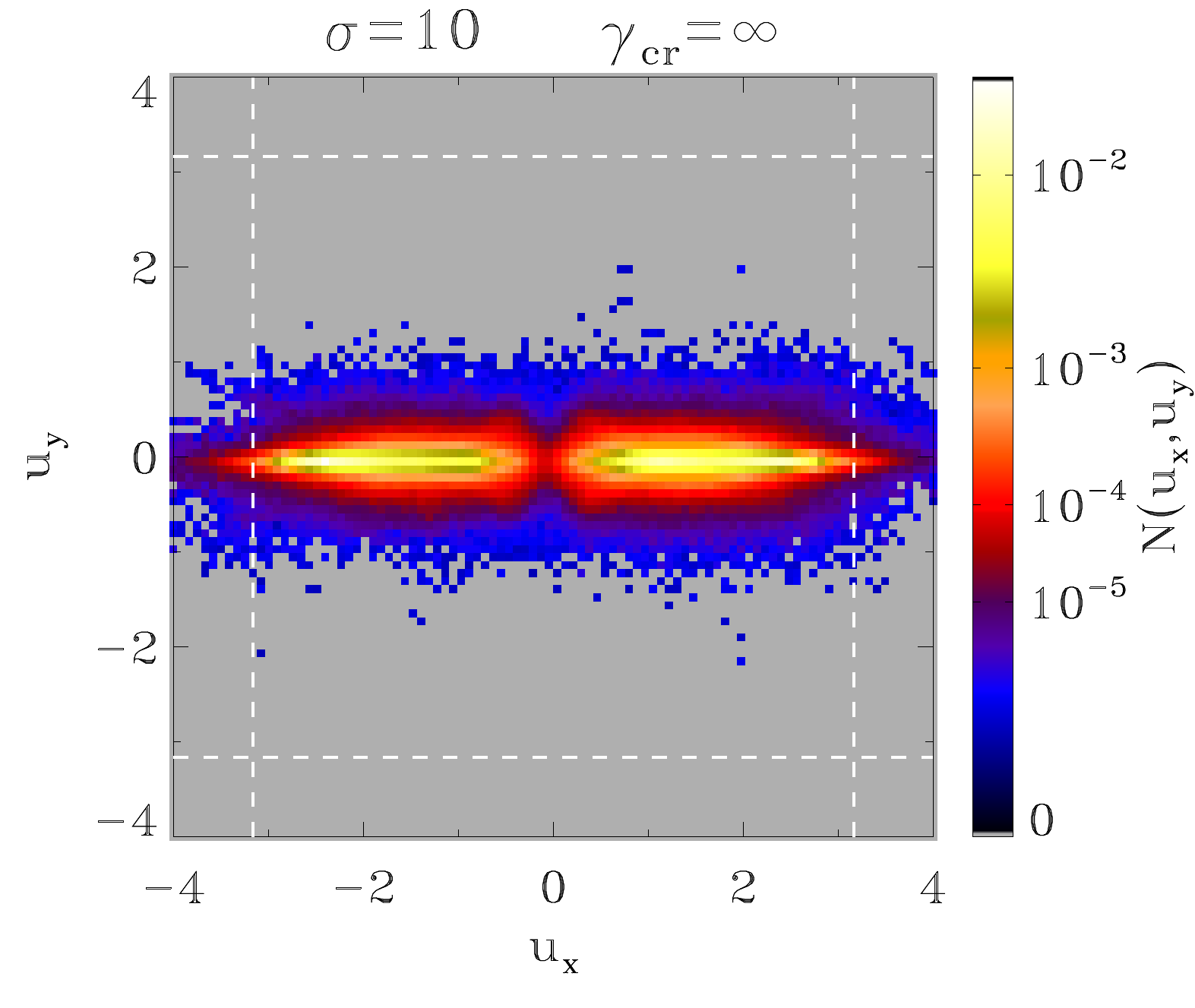}}
 \caption{Density-weighted distribution of positron bulk momenta, from the same simulations as in \fig{fluid2} ($\gammacr=16$; left panel) and  \fig{fluid1} ($\gammacr=\infty$; right panel). A similar distribution is found for the electron component (not shown in the figure). 
 The statistics of bulk four-velocities $\mathbf{u}=\gamma{\bmath \beta}$ was accumulated by averaging snapshots over the time period
 $1.5\lesssim ct/L\lesssim 5$. Dashed white lines indicate the dimensionless 4-velocity $|u_{x,y}|=\sqrt{\sigma}$ expected for the fastest outflows in relativistic reconnection.
 }
  \label{fig:xy1}
\end{figure*}
%%%%%%%%%%%%%%%%%%%%%
\begin{figure*}
\centering
\resizebox{\hsize}{!}{\includegraphics{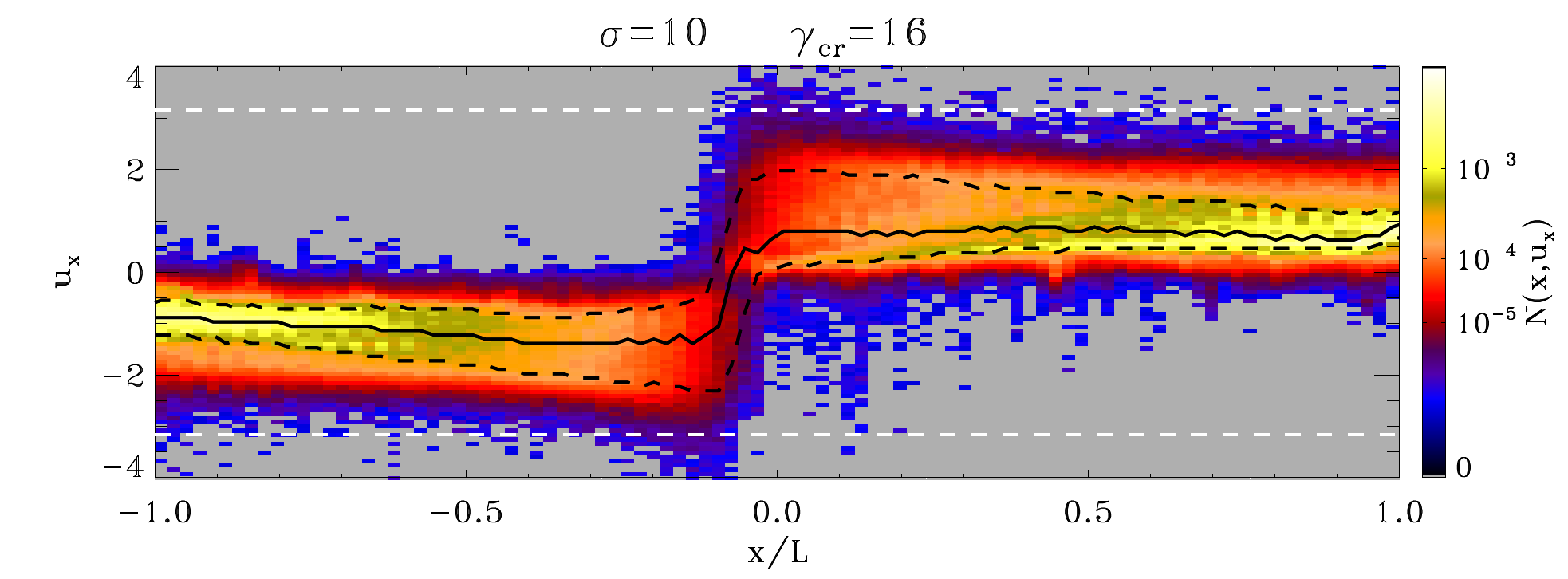}}
\resizebox{\hsize}{!}{\includegraphics{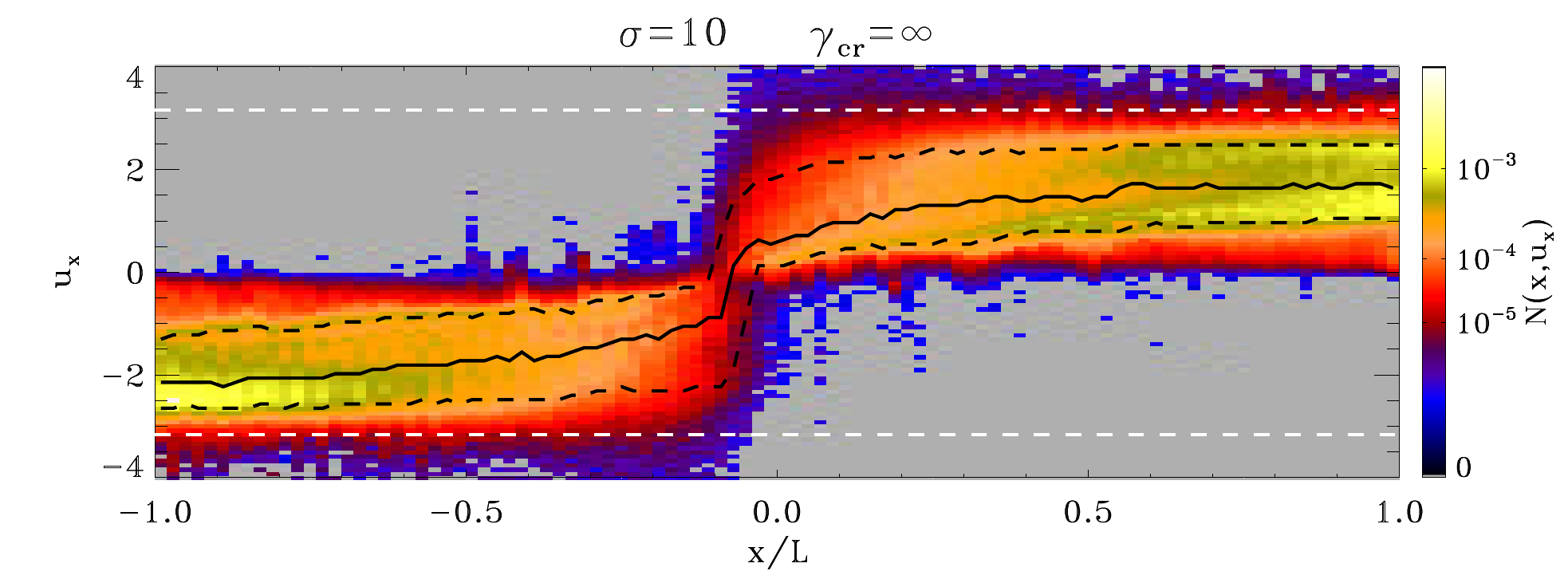}}

 \caption{
Scatter of positron bulk motions in the $x$-direction (the direction of the reconnection outflow) from the same simulations as in \fig{fluid2} ($\gammacr=16$; top) and \fig{fluid1} ($\gammacr=\infty$; bottom). The scatter results from spatial and temporal variations of the plasma flow in the reconnection region. 
The plot was obtained by averaging 70 snapshots over the extended time period $1.5\lesssim ct/L\lesssim 5$ to convey a representative quasi steady-state picture. In each snapshot, the positron bulk motion was measured using the $x$-component of the positron bulk four-velocity $u_{x}$. The measurement was made in each grid cell of the reconnection region (as defined in the text), and assigned weight proportional to the local positron density. At a fixed $x$, the result varies in time and also depends on the $y$ position of the cell. This variability generates the distribution of $u_x$ shown in the figure. The solid black curve shows the median value of  $u_x$ as a function $x$ (the mean expectation of $u_x$ gives nearly the same curve),
and the dashed black curves indicate the $10\%$ and 90\% percentiles of the $u_x$ distribution.  The horizontal dashed white lines indicate $\pm\sqrt{\sigma}$, the value of 4-velocity expected for the fastest reconnection outflows.
}
 \label{fig:pspace2}
\end{figure*}
%%%%%%%%%%%%%%%%%%%%%

%%%%%%%%%%%%%%%%%%%%%%%

\subsection{Flow structure}\label{sec:flow}

The overall structure of the reconnection flow with strong IC losses $(\gammacr=16$) is presented in \fig{fluid2}. It should be compared with \fig{fluid1}, which shows the simulation without IC losses, with otherwise identical parameters. One evident difference 
is the low internal energy in the radiative case. This is expected, as heat is lost to IC 
radiation.

The magnetic field structure displays a striking similarity in the two runs. 
The opening angle of the magnetic field lines in the inflow region is closely related to the inflow speed
\citep{lyubarsky_05}, and
so the similarity 
of these angles in \fig{fluid2} and \fig{fluid1} indicates that the reconnection rate weakly depends on $\gammacr$. We find that the time-averaged reconnection rate is $\eta_{\rm rec}=v_{\rm rec}/c\approx 0.122$ in the non-radiative simulation and $\eta_{\rm rec}/c\approx 0.135$ in the radiative simulation. The weak dependence of $\eta_{\rm rec}$ on the degree of radiative losses is further demonstrated in \sect{scan}.

In both radiative and non-radiative simulations,
the reconnection layer fragments into a chain of plasmoids of various sizes, which appear as strongly magnetized structures.  At the centers of plasmoids, the parameter $\epsilon_{\rm B}\equiv B^2/8\pi n_0 m_e c^2$ (the ratio of magnetic energy density to upstream rest mass energy density) reaches $\sim 100$, regardless of the level of radiative losses. This should be compared with its initial value in the inflow region, $\epsilon_{\rm B}\approx\sigma/2=5$. The magnetic field $B$ in the plasmoid core is 
compressed by a factor of $B/B_0\sim 4$. When averaged over the plasmoid volume, the mean magnetic energy density in the plasmoid rest frame is roughly twice that in the upstream (as shown by \citet{sironi_16} for cases with zero guide field $B_g=0$).
The similarity of the magnetic field structure in \fig{fluid2} and \fig{fluid1} demonstrates that the plasmoid chain formation is controlled by magnetic stresses, and plasma pressure is not crucial even in the hot (non-radiative) regime. 

Plasma density structure, on the other hand, exhibits significant differences between the two runs (compare panels (b) of \fig{fluid2} and  \fig{fluid1}). 
In the radiative simulation, the plasmoids are strongly asymmetric: the plasma trapped in magnetic islands 
tends to pile up 
in the back of the moving island, leaving a cavity at the front 
(see, e.g., the plasmoids at $x/L\sim -0.8$ and at $x/L\sim 0.9$  in \fig{fluid2}(b)).  
This effect is caused by Compton drag on the plasma particles 
while most of the island inertia is in the magnetic field. Since plasma can slide along the magnetic field lines, Compton drag partially succeeds in resisting the plasma motion and keeps it in the back of the island.
This is analogous to a fast-moving car with open top and without a wind shield, exposing passengers to air resistance. The air drag (Compton drag) pushes the passengers (plasma particles) toward the back of the car (magnetic island).  
 
The biggest change in the radiative case is that most of the power released by reconnection is radiated away rather than stored in the plasma. In particular, almost all internal energy received by the plasma is lost.
As a result, the
IC emission is generated predominantly by the  bulk motions of cooled plasmoids (especially the large ones). The approximate equality of the total and bulk IC losses is demonstrated by the nearly identical colors of large plasmoids in panels (c) and (d) of \fig{fluid2} (see e.g., the plasmoid at $x/L\sim -0.8$).
This confirms the picture described in B17: most of the reconnection energy goes into work by magnetic tension forces to move cold plasmoids against Compton drag. As a result, most of the released energy converts directly to radiation through bulk Comptonization.

A smaller fraction of the dissipated reconnection power goes into impulsive acceleration of individual particles, followed by their quick cooling.\footnote{\rev{In the following, by ``dissipated reconnection power'' we indicate the energy per unit time transferred by reconnection from the fields to the particles. This is a fraction $\sim 50\%$ of the Poynting flux carried towards the reconnection layer, since half of the incoming  energy stays locked in the magnetic field \citep{sironi_15}.} } Individual particles are energized outside the plasmoids, in particular at X-points or in fast outflows emanating from the X-points. This is clearly observed in the thin regions of the reconnection layer, where the freshly energized plasma has not cooled yet (see e.g. the thin red layer at $-0.2\lesssim x/L\lesssim 0.1$ in \fig{fluid2}(e)). 
The same happens at the interfaces between merging plasmoids where transient ``secondary'' current sheets are formed perpendicular to the main current sheet (see the vertical red layer at $x/L\sim -0.6$ or $x/L\sim 0.3$). Here particles are actively heated and accelerated just like in the primary current sheet.

\subsection{
Bulk motions in the reconnection layer
}\label{sec:plasm}

We now describe in more detail the plasma bulk motions in the radiative reconnection layer.
We measure the bulk motion (for each species, electron and positron) as follows. 
For each cell, we compute the average particle velocity including all the particles in the neighboring $5\times5$ cells: $\bmath{\beta}=\langle\bmath{\beta}_e\rangle$. 
In the frame moving with velocity $\bmath{\beta}$, 
the plasma stress-energy tensor has a vanishing energy flux \citep[e.g.,][]{rowan_19}, and we define this frame as the comoving frame of the plasma (see e.g., \citealt{zhdankin_18a} for alternative definitions). The corresponding bulk Lorentz factor is given by $\gammab=(1-\betab^2)^{-1/2}$.

We have measured the statistics of bulk motions in the reconnection layer using the positron component as a proxy. We include only particles inside the reconnection layer. This layer is defined based on the degree of mixing between the two opposite vertical inflows toward the midplane $y=0$
(see e.g., \citet{rowan_17} and \citet{ball_18}). Particles coming from above ($y>0$) and below ($y<0$) are tagged as two different populations, and the reconnection layer is defined as the region where both populations contribute at least $1$\% to the local plasma density.  
The results shown below are nearly insensitive to the exact value of the mixing threshold, as long as it is significantly above zero. 

There is some $y$-component of bulk motions, which is produced by vertical secondary reconnection layers between merging plasmoids.\footnote{Faster bulk motions in the $y$ direction are attained for the radiative case $\gammacr=16$ (left panel) than for the uncooled case $\gammacr=\infty$ (right panel). This is related to the density cavities at the front of cooled plasmoids, where the local magnetization reaches high values (due to the very low density), resulting in fast reconnection outflows in between merging plasmoids.} However, the dominant component of bulk motions is along the $x$-axis 
(\fig{xy1}), and below we focus on the $x$-component. 
It strongly varies in space and time. 
The scatter of these variations may be described by an effective temperature, which 
 influences the hard X-rays spectrum produced by the reconnection layer through Comptonization (B17).
%-----------

\fig{pspace2} presents the distribution of $u_x=\gammab\betabx$ measured at each vertical slice of the reconnection layer during an extended time period $1.5\lesssim ct/L\lesssim 5$. The result depends on the slice position $x$. We observe that the horizontal motions have a regular component
(represented by the median or the mean expectation of $u_x(x)$)
and a strong stochastic component. 
The regular component represents the plasma outflow away from $x=0$ (the center of the layer). By symmetry, the outflow speed is zero at $x=0$ and increases toward the edges $x=\pm L$, approaching $|u_x|\approx 2$ in the non-radiative model and $|u_x|\approx 1$ in the radiative model. This regular component is below $|u_x|\approx \sqrt{\sigma}\approx 3$ predicted by analytical models \citep{lyubarsky_05}. The strong reduction of the mean outflow speed in radiative reconnection is the result of Compton drag.

Superimposed on the average outflow are substantial stochastic motions. They are particularly strong for small plasmoids, which are capable of approaching the maximum $|u_x|\sim\sqrt{\sigma}$ even for the radiative case. About 10\% of the time, $|u_x|$ exceeds 2.  A more typical variation is comparable to the mean expectation value.
The stochastic bulk motions are caused by continuing tearing instabilities forming new plasmoids, and their subsequent mergers. Large old plasmoids tend to attract and accrete small young plasmoids, and so
some young plasmoids are pulled back by a large plasmoid behind them.
As a result, a small fraction of plasma motions are directed toward the center $x=0$, opposite to the mean expectation.

The observed distributions of four-velocity $u_x$ do not, however, provide a useful physical picture of the Compton drag effect on the reconnection chain. They are ignorant of the fact that most of reconnection energy resides in distinct, isolated plasmoids of various sizes $w$, as clearly seen in panel (a) of \fig{fluid2}.
Therefore, we extend our analysis of the reconnection layer by examining the motions of individual plasmoids. Similar to 
\citet{sironi_16},
we identify the individual magnetic islands using the magnetic vector potential.
The islands are typically found to have a quasi-elliptical shape, with a transverse size $w$ and a comoving length of $\sim 3w/2$.
The plasmoid four-velocity $\upl$ is defined as the bulk $u_x$ of the center of  the magnetic island (the peak of the vector potential).

The motion of a plasmoid is governed by two competing forces:
it is accelerated by magnetic tension and decelerated by Compton drag. The outcome of this competition depends on the plasmoid size.
B17 argued that the magnetic tension forces
in radiative reconnection should be approximately the same as in non-radiative reconnection, and 
estimated the chain properties
based on this conjecture. The basic idea of this method is as follows.
In the absence of Compton drag, the measured plasmoid motions directly reflect the magnetic forces $f_B$ accelerating them. 
Thus the non-radiative simulation gives $f_B(w,t^\prime_{\rm age})$ acting on plasmoids of various sizes $w$ and proper ages $t^\prime_{\rm age}$.\footnote{Hereafter primed quantities refer to the plasmoid comoving frame. The transverse size $w$ is invariant under Lorentz boosts along the direction of plasmoid motion.}
If $f_B(w,t^\prime_{\rm age})$ indeed remains the same for radiative reconnection, one can simply add Compton drag $f_{\rm drag}$ and evaluate the dynamics of plasmoids subjected to two known forces, $f_B$ and $f_{\rm drag}$.
Below we test 
this method against the results of our numerical simulations. 

Previous analysis of non-radiative reconnection simulations demonstrated 
a close relation between $t_{\rm age}^\prime$, $w$, and the plasmoid four-velocity $\upl$ \citep{sironi_16}. 
This relation is described by the empirical formula,
\be\label{eq:nocool}
\frac{\upl}{\sqrt{\sigma}} \approx  \tanh\left(\frac{\upl}{\sqrt{\sigma}}\,\frac{\etarec ct_{\rm age}^\prime}{w}\right),
 \qquad \upl=\gamma\beta\approx \gamma\beta_x.
\ee
It shows that plasmoids with $\etarec ct_{\rm age}^\prime/w>1$ approach the maximum, saturated four-velocity $\upl\approx\sqrt{\sigma}$. 
It also shows that before the saturation occurs, i.e. when $\upl$ is well below $\sqrt{\sigma}$, plasmoids grow proportionally to their proper ages, $w\approx \etarec c t_{\rm age}^\prime$, and the growth slows down after the saturation of $\upl$.

A large part of the plasmoid momentum is carried by the magnetic field (even in a hot, non-radiative chain). After averaging over the plasmoid volume, its  momentum density is given by
\be
  p_{\rm pl}\sim \frac{2U_B\,\betab}{c}\sim \frac{B_0^2\gammab^2\betab}{2\pi c},
\ee 
where we used the relation $U_B\sim \gammab^2B_0^2/4\pi$ measured by \citet{sironi_16}.
Plasmoids acquire this momentum during time $t_{\rm age}=\gammab t^\prime_{\rm age}$. This implies an average force per unit volume
$f_B\sim p_{\rm pl}/t_{\rm age}$ and gives
\be
\label{eq:fB}
  f_B\sim \frac{B_0^2\upl}{2\pi c t^\prime_{\rm age}}.
\ee
Equations~(\ref{eq:nocool}) and (\ref{eq:fB}) allow one to exclude $t^\prime_{\rm age}$ and find $f_B$ acting on a plasmoid with given $w$ and $\upl$.

We can now compare $f_B$ with the Compton drag force $f_{\rm drag}$ (B17),
\be
  f_{\rm drag}\approx \frac{4}{3}\,\gammab^2 U_{\rm rad} \sigma_{\rm T} \gh^2 \betab \, n_{\rm pl}.  
\ee
Here $n_{\rm pl}$ is the mean density of the plasmoid, and $\gh$ is the characteristic Lorentz factor of particles measured in the plasmoid rest frame; $\gh-1$ is a measure of heat.
For most plasmoids (except the smallest ones), the drag timescale $t_{\rm drag}=p_{\rm pl}/f_{\rm drag}$ satisfies the following conditions (see B17),
\be
   t_{\rm IC}<t_{\rm drag}<t_{\rm age}.
\ee
This implies that the plasmoids are efficiently cooled to $\gh\approx 1$ (we check the accuracy of this condition for our simulations below). 
Then one finds
\be
\label{eq:ratio}
   \frac{f_{\rm drag}}{f_B}\sim \frac{8\pi}{3} \,\frac{\sigma_{\rm T}U_{\rm rad}\gammab^2 n_{\rm pl}^\prime c t_{\rm age}^\prime}{B_0^2}
   \approx \frac{\etarec}{2\sqrt{\sigma}}\,\frac{n_{\rm pl}^\prime}{n_0}\,\frac{\gammab^2}{\gammacr^2}\,\omega_{\rm p} t_{\rm age}^\prime,
\ee
where $n_{\rm pl}^\prime\approx n_{\rm pl}/\gammab$ is the mean proper density of the plasmoid.

The main predicted effect of Compton drag is that the plasmoid acceleration should cease when $f_{\rm drag}=f_B$, and so plasmoids should be prevented from entering the regime where $f_{\rm drag}>f_B$. 
The boundary of this ``avoidance zone'' on the $w$-$u$ plane can be found by setting $f_{\rm drag}/f_B=1$ in \eq{ratio}, then expressing $t_{\rm age}^\prime$ from this equation, and substituting the result into \eq{nocool}. This gives,
\be
\label{eq:umax}
   \upl=\sqrt{\sigma}\tanh\left(\frac{2n_0}{n_{\rm pl}^\prime} \frac{\comp}{w}\frac{\gammacr^2\,\beta}{
      \gamma}
  \right). 
\ee
Using the compression factor $n_{\rm pl}^\prime/n_0\approx 6$ measured in the simulation with $\gammacr=16$ and $L=3360\comp$, one can solve \eq{umax} for $\upl$ for plasmoids of a given size $w/L$. The solution represents the expected boundary of the avoidance zone on the $w$-$u$ plane. Plasmoids are expected to congregate near this boundary, and we can quantitatively test this expectation by measuring the plasmoid speeds in the simulation.

%%%%%%%%%%%%%%%%%
\begin{figure}
\centering
\resizebox{1\hsize}{!}{\includegraphics{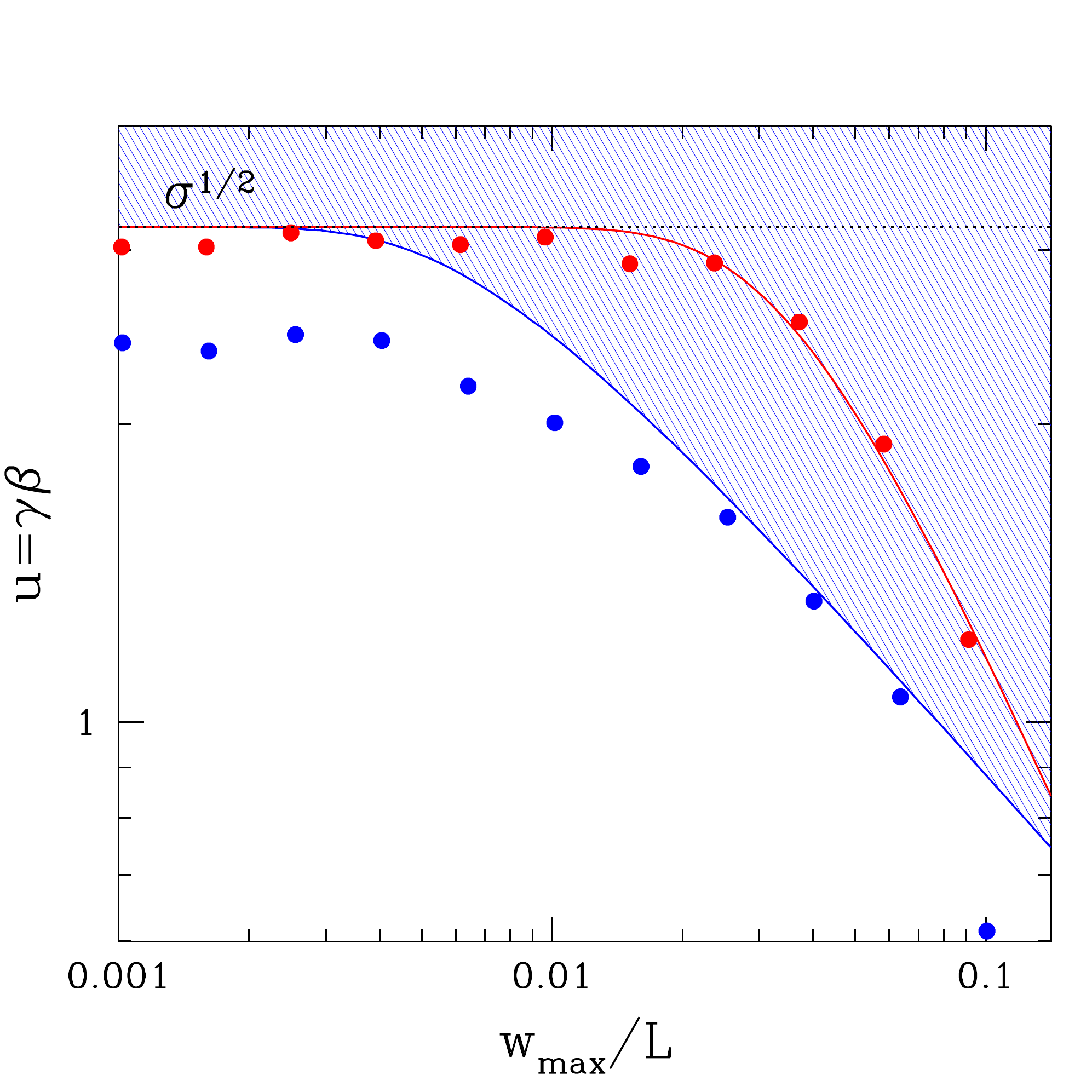}}
 \caption{
The outcome of plasmoid acceleration as a function of its size. Plasmoid four-velocities $u$ were
 measured when the plasmoid reached its maximum size $w_{\max}$, i.e. typically at the end of its life in the simulation --- either at a merger or when the plasmoid exits the computational box. 
Blue and red points show the results in the radiative ($\gammacr=16$) and non-radiative ($\gammacr=\infty$) simulation, respectively. The red curve shows the analytical formula in \eq{bulk2}. The avoidance zone expected in the radiative regime and described by \eq{umax} is shaded in blue.   
}
 \label{fig:mommax}
\end{figure}
%%%%%%%%%%%%%%%%%

\fig{mommax} shows the 4-velocity $\upl$ of plasmoids at the end of their lives, before they exit the computational box or merge with another plasmoid. More exactly, the measurement of $\upl$ was taken when the plasmoid reached its maximum transverse size $w_{\max}$, and the results are presented as a function of $w_{\max}$. \fig{mommax} shows the results for  non-radiative ($\gammacr=\infty$) and radiative ($\gammacr=16$) reconnection. 

In the non-radiative regime, the measurements are consistent with \eq{nocool} when we substitute\footnote{
This substitution approximately corresponds to measuring the plasmoid speeds at the exits of the computational box ($x=\pm L$). It somewhat overestimates the average $t_{\rm age}^\prime(w_{\max})$, as many plasmoids end their lives in mergers, at a smaller $t_{\rm age}^\prime$. Therefore, \eq{bulk2} somewhat overestimates $u(w_{\max})$.}
 $\gammab\betab c t^\prime_{\rm age}\sim L$, which gives 
\be\label{eq:bulk2}
  \upl\approx \sqrt{\sigma} \tanh\left(\frac{\etarec}{\sqrt{\sigma}}\frac{L}{w_{\rm max}}\right)~~.
\ee
In agreement with previous simulations \citep{sironi_16}, we observe the saturation of $\upl\approx \sqrt{\sigma}$ for small plasmoids, with $w_{\max}/L\lesssim (\eta_{\rm rec}/\sqrt{\sigma})\approx 0.03 $. 

A similar saturation should hold in the radiative regime, however now there is an additional limiting factor --- Compton drag. For radiative reconnection, we compare the measured $\upl(w_{\max})$ with \eq{umax}, and observe a good agreement in the range $0.003< w_{\rm max}/L<0.1$. This quantitative test confirms the conjecture that magnetic tension forces are approximately the same in radiative and non-radiative regimes. Deviations are observed for the smallest plasmoids and caused by $\gh>1$.

The deviation of $\gh$ from unity is small for most plasmoids in radiative reconnection, because
radiative losses dramatically reduce the plasma internal energy.
The cooling timescale measured in the plasmoid rest frame is given by (B17),
\be\label{eq:tic}
t^\prime_{\rm IC}(\gh)=\frac{3m_ec}{4\sigma_{\rm T}U_{\rm rad}^\prime\gh}
   \approx \frac{\gammacr^2}{\eta_{\rm rec} \sqrt{\sigma}\gammab^2\gh}\,\omega_{\rm p}^{-1}~,
\ee
where $U_{\rm rad}^\prime\sim\gammab^{2}U_{\rm rad}$ from the  transformation of $U_{\rm rad}$ to the plasmoid frame.
Cooling occurs faster than Compton drag. The drag effect during the cooling timescale $t_{\rm IC}=\gammab t_{\rm IC}^\prime$ may be expressed as 
\be
  \frac{f_{\rm drag} t_{\rm IC}}{p_{\rm pl}}\sim \frac{1}{\sigma_{\rm pl}}, 
  \qquad \sigma_{\rm pl}=\frac{2U_B}{U_{\rm pl}}=\frac{B^2}{4\pi n_{\rm pl}\gammab\gh m_ec^2}.
\ee
Non-radiative reconnection with $\sigma\gg 1$ would create plasmoids with $\sigma_{\rm pl}\sim 1$. By contrast, radiative relativistic reconnection gives $\sigma_{\rm pl}\gg 1$  for almost all plasmoids except the young/small ones that have not cooled yet but have already reached $\gammab\approx \sqrt{\sigma}$. 
The small plasmoids must cool to $\gh\approx 1$ when their proper ages reach $t_{\rm IC}^\prime$ evaluated with $\gh\sim 1$ and $\gammab\sim\sqrt{\sigma}$. This condition defines the ``cooling age.'' The corresponding characteristic ``cooling size'' of plasmoids is given by
\be
\label{eq:wc}
  w_c\sim \etarec ct_{\rm IC}^\prime\sim \frac{\gammacr^2}{\sigma^{3/2}}\,\frac{c}{\omega_{\rm p}}.
\ee
 In particular, in our fiducial model with $\gammacr=16$ and $L/(\comp)\sim 3360$, we find $w_c/L\sim 0.003$. Thus, all plasmoids with $w/L\gg 0.003$ should be cooled to $\gh\sim 1$. 
 
\begin{figure}
\centering
\resizebox{\hsize}{!}{\includegraphics{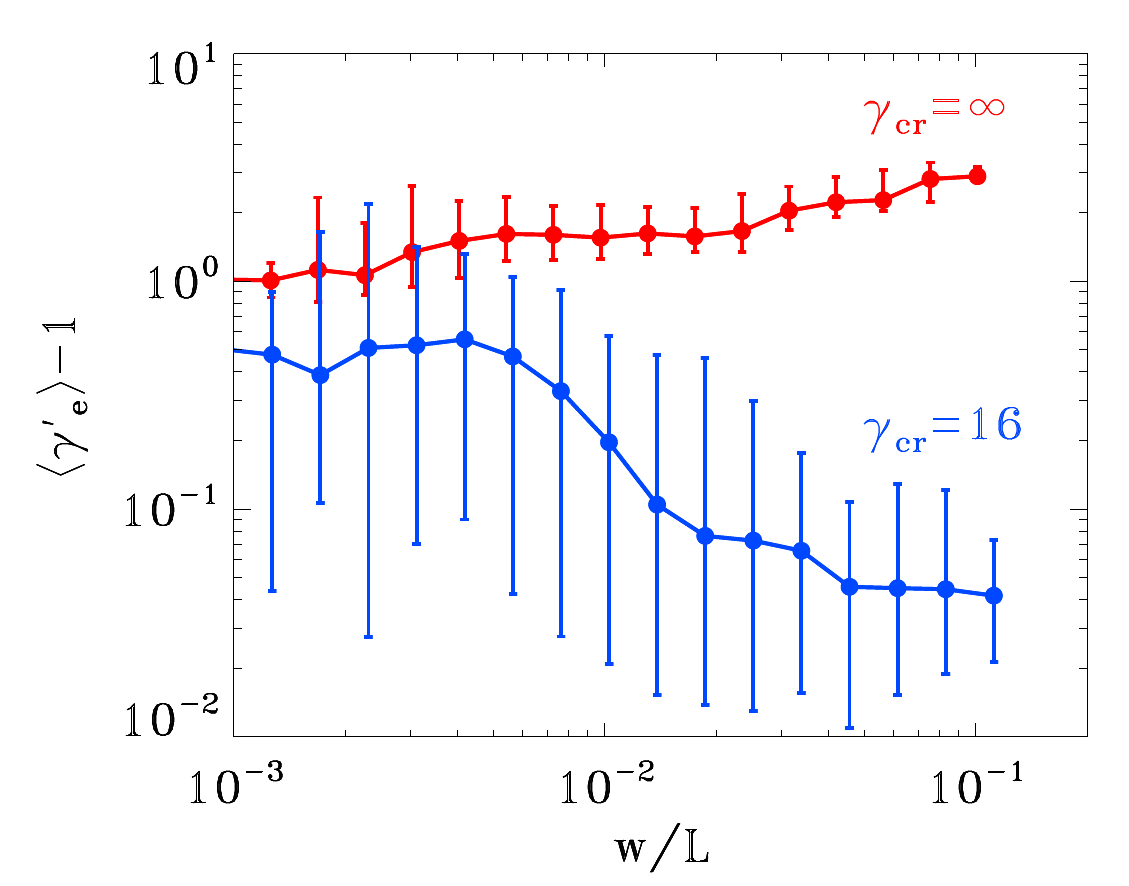}}
 \caption{Mean internal energy per particle 
(in units of $m_ec^2$)
 in plasmoids, as a function of the plasmoid transverse size $w$,
in the radiative (blue) and non-radiative (red) simulations.
 The filled circles connected by the lines show the median values, whereas the error bars show the 10\% and 90\% percentiles.}
 \label{fig:engadi}
\end{figure}

This analytical estimate 
is confirmed in \fig{engadi}, where we plot the mean internal energy per particle in plasmoids as a function of the plasmoid transverse size. In the radiative case (blue curve), 
we observe
$\langle\gamma_e'\rangle\approx 1.5$ for plasmoids 
with $w/L\lesssim  0.003$. 
This explains why the bulk motions of small plasmoids (blue points in \fig{mommax})
stay somewhat below the prediction of \eq{umax}, which assumed $\gh=1$.
For plasmoids with $w/L\gtrsim  0.003$, $\langle\gh\rangle$ drops toward unity, in agreement with the 
estimate in \eq{wc}.
We have also verified that $w_c$ remains consistent with  \eq{wc} in a larger simulation (with doubled $L$), i.e. $w_c/(\!\!\unit{\comp})$ remained unchanged.

In contrast, for non-radiative reconnection we always find $\langle\gamma_e'\rangle\gtrsim 2$, with a 
tendency for hotter particles in larger plasmoids (red curve in \fig{engadi}). 

%###############################################

\begin{figure}
\centering
\resizebox{1\hsize}{!}{\includegraphics{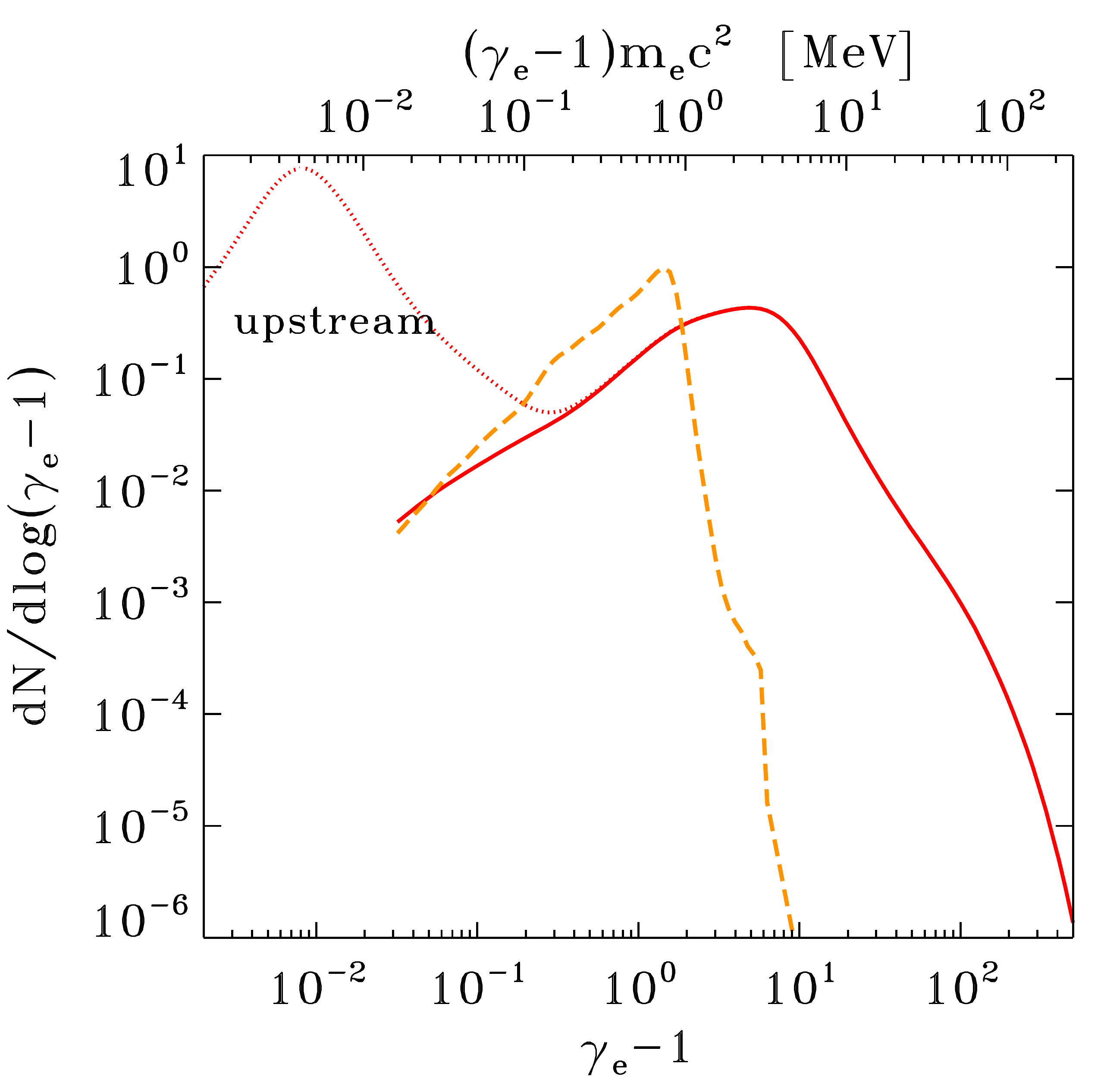}}
 \caption{Particle energy 
 distribution in the non-radiative simulation 
 ($\gammacr=\infty$), time-averaged in the interval $1.5\lesssim ct/L\lesssim 5$. The solid red curve shows the particle distribution $dN/d\log(\gamma_e-1)$ in the reconnection region (as defined in \sect{plasm}).
The dotted red curve
shows
the distribution 
in the 
entire computational 
box, including the cold inflow region where particles move with speed $v\approx \eta_{\rm rec} c$ and form the low-energy peak at $\gamma_e-1\sim 10^{-2}$.
The dashed orange curve shows the particle distribution in the reconnection region when accounting only for {\it bulk} kinetic energy (i.e., we plot $dN/d\log(\gammab-1)$ as a function of $\gammab-1$). 
All three distributions 
are normalized 
to the total number of particles in the reconnection region.
}
 \label{fig:spec}
\end{figure}

\begin{figure}
\centering
\resizebox{1\hsize}{!}{\includegraphics{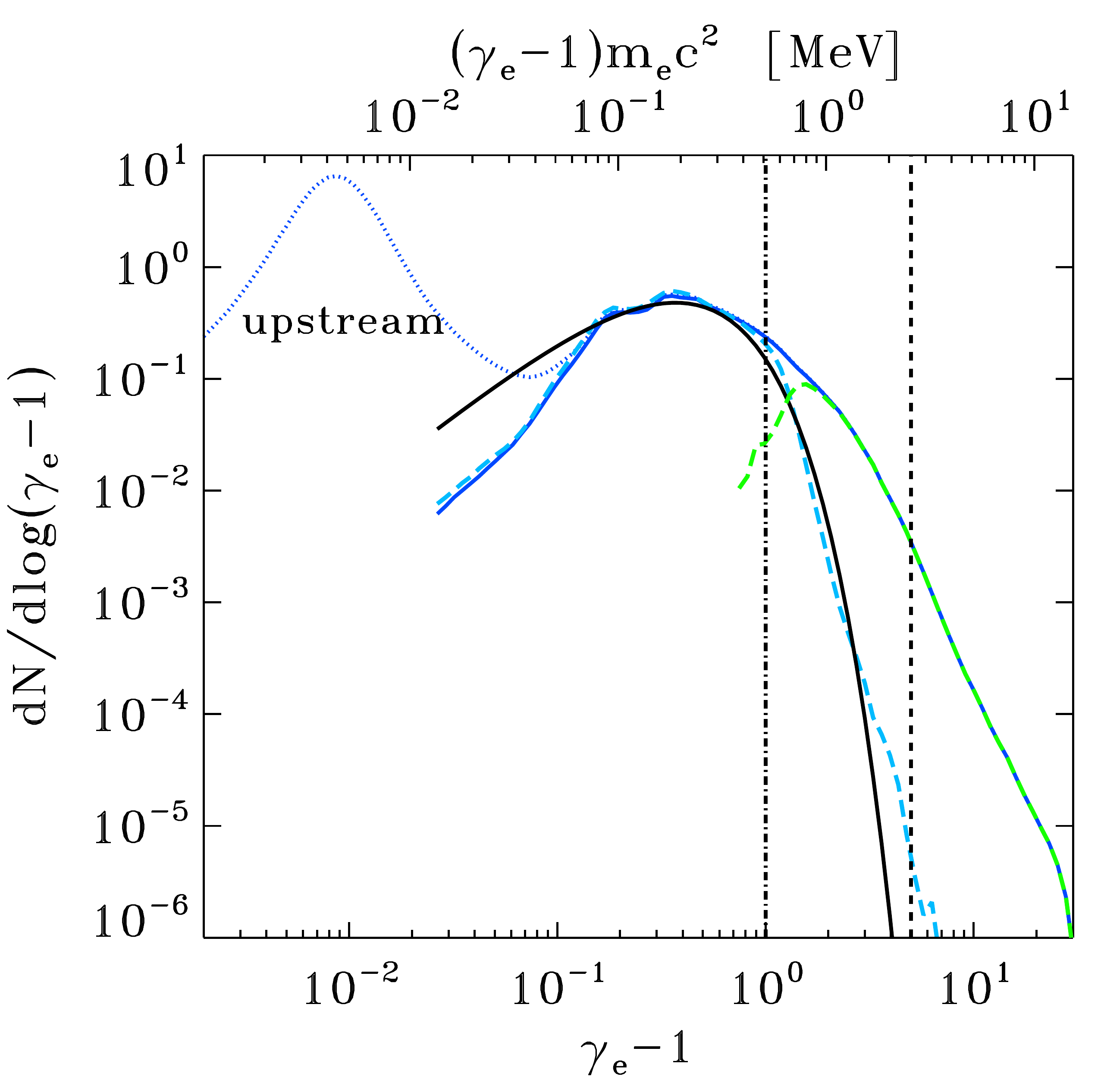}}
 \caption{
 Particle 
 distribution in the radiative simulation
($\gammacr=16$), time-averaged in the interval $1.5\lesssim ct/L\lesssim 5$. Similar to \fig{spec}, we show $dN/d\log(\gamma_e-1)$ in the reconnection region (solid blue) and in the entire computational box (dashed blue).
The dashed cyan curve shows the particle distribution in the reconnection region when accounting only for {\it bulk} kinetic energy (i.e., we plot $dN/d\log(\gammab-1)$ as a function of $\gammab-1$).
For comparison, the solid black curve shows a Maxwellian distribution with temperature $kT_{\rm b}=100$ keV.
The dashed green curve is the difference between the solid blue and the dashed cyan curves; it represents the part of the energy distribution that is not accounted for by bulk motions.
The two vertical dashed lines approximately show the divison into three energy regions where different acceleration processes dominate: bulk acceleration by magnetic tension forces ($\game\lesssim 2$), particle pick-up by outflows from X-points ($2<\game<6$), and X-point acceleration ($\game\gtrsim 6$).  
}
 \label{fig:specfit}
\end{figure}

%%%%%%%%%%%%%%%
\subsection{
High-energy particles
}\label{sec:spect}

In non-radiative reconnection, a large fraction of the released magnetic energy becomes stored in relativistic, $\gh\gg 1$, quasi-isotropic internal motions of particles in the plasma rest frame.
This fact is demonstrated in \fig{spec}. The figure compares the particle distributions over the bulk motion energy $(\gammab-1)m_ec^2$ and the actual particle energy $(\game-1)m_ec^2$. 
One can see that
the distribution $dN/d\log(\game-1)$ is strongly shifted toward high energies compared with $dN/d\log(\gammab-1)$. This reflects the fact that non-radiative reconnection heats the plasma to $\gh\gg 1$, and so particles gain $\game\gg \gammab$ (note that $\game\sim\gammab\gh$). 
The hot distribution peaks at $\game\simeq6\sim 0.5\sigma$. 
The broad bump at $\game\sim0.5\sigma$ is dominated by plasma between plasmoids, in particular by particles in the outflows from X-points (they are energized by the pick-up process discussed in detail below). The slope $p\equiv d\log N/d\log\game$ on the high-energy side of this bump is steep, making the appearance of a power-law tail with $p\sim 3.5$.\footnote{
     This high-energy slope is  
significantly steeper than $p\sim 2$ usually reported by PIC simulations of relativistic non-radiative reconnection
\citep[e.g.,][]{ss_14,guo_14,werner_16}. As discussed in \citet{sironi_16}, this difference comes from the choice of boundary conditions at $y=\pm L$ where the plasmoids exit the box.
Our simulation employed the outflow boundary conditions, so the growing plasmoids disappear as they exit the box. This leads to a relatively large contribution from plasma between the plasmoids.
In contrast, in simulations with periodic boundaries, the plasmoids stay in the box and keep growing. Then the particle distribution becomes dominated by 
the largest plasmoids, filled with a broad flat distribution. 
\citet{sironi_16} demonstrated that, regardless of the chosen boundary conditions, the particle spectrum inside large plasmoids 
has the slope
$p\sim 2$, and 
with open boundary conditions the inter-plasmoid bump around $\game\sim0.5\sigma$ emerges above the plasmoid spectrum. 
}

Strong IC losses change the particle distribution (\fig{specfit}). There are three main changes. First, nonthermal particle acceleration to $\gamma_e\gg\sigma$ is 
suppressed (see also \citealt{werner_19}). Second, 
the peak of $dN/d\log(\game-1)$ is shifted from $\game\sim0.5\sigma$ to a much lower, mildly relativistic value $\game\sim 1.3$. This change reflects the fact that the plasma loses most of its energy to radiation. Third, the distribution shape at $\game<2$ becomes controlled by bulk motions, i.e. the particles populating the peak are cold, with $\gh\approx 1$ and $\game\approx \gammab$.  

The bulk-motion distribution $dN/d\log(\gammab-1)$ steeply declines above $\gamma\sim 2$ 
and its shape can be
approximately
described as Maxwellian, with an effective ``bulk temperature'' $k\Tb\approx 100$~keV.
At Lorentz factors $\game\sim 2$ (kinetic energies $\sim m_ec^2= 511$~keV), the distribution $dN/d\log(\game-1)$ strongly deviates from $dN/d\log(\gammab-1)$ and forms an extended tail with a slope $p\sim -4.3$. 
Thus, the 
distribution can be rather cleanly separated into two distinct components: the main 100-keV peak shaped by the bulk motions of cold plasma 
and the high-energy component dominated by individual motions of nonthermal particles. 
The high-energy component extends from $\game\sim 2$ to $\game$ of a few $\sigma$.\footnote{The high-energy tail extends up to $\gamma_e\sim 30$, contrary to the anticipated cutoff at $\gamma_e\sim\gammacr=16$ (see \eq{gammacr1}). The cutoff is pushed to higher $\gamma_e$ because the reconnection layers between merging plasmoids have a stronger $B>B_0$ and hence a stronger $E\approx 0.1B$, capable of pushing particles to higher energies.}

The cooling time of energetic particles is short, and so the observed extension of 
$dN/d\log(\game-1)$
from 500~keV to $\sim 10$~MeV must be sustained by a process of nonthermal particle injection in the reconnection layer. Indeed, we observe that nearly impulsive acceleration of particles operates at the X-points and  outflows from the X-points, as described below.

The high-energy component receives a significant fraction of the total power dissipated by magnetic reconnection. We can measure this fraction $\fHE$ using the fact that the received power is immediately radiated away, with the emission rate proportional to $\game^2\beta_e^2$. We have computed the ratio of the power emitted by the high-energy particles ($\game>2$) and the net power emitted by all particles in the reconnection region. This ratio is $\fHE\approx 35$\% for our fiducial model shown in \fig{specfit} ($\gammacr=16$ and $L/(\comp)\simeq 3360$). We have also explored how $\fHE$ depends on the simulation parameters, and found that it decreases to $\fHE\approx 27$\% for a larger computational box $L/(\comp)\simeq 6720$. This decrease is moderate, indicating convergence with increasing box size.
We obtained similar values for $\fHE$ in models with yet stronger IC cooling (lower $\gammacr$). In particular, in  simulations with $\gammacr=11.3$ we find $\fHE\approx 26$\% for $L/(\comp)\simeq 3360$ and $\fHE\approx 22$\% for $L/(\comp)\simeq 6720$. We conclude that $\fHE\sim 20$\% in the strongly radiative regime.

The injection of high-energy particles is important for the radiation spectrum emitted by the reconnection layer. Therefore, we further investigated the mechanism of nearly impulsive acceleration observed in our simulations.  We have analyzed the histories of individual particles and identified two mechanisms 
generating the high-energy component.
(1) The distribution at 
$\gamma_e\gtrsim\sigma=10$ is dominated by particles ejected from the X-points by the accelerating electric field $E\sim 0.1B_0$. This mechanism was studied in detail in previous works \citep[e.g.,][]{zenitani_01}.
(2) At intermediate energies, $2\lesssim \game\lesssim 6$, the distribution is dominated by particles located between the X-points and the neighboring plasmoids. We call 
these
regions  ``unstructured outflows'' from the X-points. We found that the energetic particles appearing in the outflows are injected by a special ``pick-up'' process described below.

When two opposite horizontal magnetic field lines (above and below the reconnection layer) become connected at the X-point, they form a single field line with a cusp. Magnetic tension causes the cusp to snap horizontally toward the neighboring plasmoid. The snapping motion occurs with an increasing Lorentz factor $\gamma$, and the moving cusp ``picks up'' the plasma particles on the field line. This process resembles collision of a moving wall with a static particle, which gives the particle $\gamma_e\sim \gamma^2$ in the lab frame ($\game^\prime\sim \gamma$ in the wall rest frame).
The fastest unstructured outflows have $\gammab\approx\sqrt{\sigma}$ and can energize particles up to $\gamma_{\max}\sim \sigma$. In reality, this is a 
conservative
upper limit, since most of the unstructured outflows move slower, especially in the presence of Compton drag. As a result, the kicks of particles by unstructured outflows dominate the observed distribution only up to $\gamma_e\sim 6<\sigma=10$. 

The picked-up particles start to gyrate about $\bmath{B}$ with Lorentz factor $\game^\prime\approx \gamma$ in the rest frame of the moving cusp (note that 
the magnetic field at the cusp point is perpendicular to its velocity, $B_x=0$).
The gyrating particle possesses a significant oscillating $z$-momentum in the lab frame, in addition to the $x$-momentum associated with the cusp motion. This is a distinct feature of particles energized by pick-up,\footnote{In the simple case with $B_g=0$, the electromagnetic fields in the unstructured outflow at $y=0$ are dominated by $E_z$ (the reconnection electric field) and  $B_y$ (the reconnected magnetic field), and the drift speed $\bmath{E}\times\bmath{B}/B^2$ can reach $|E_z/B_y|\sim v_A/c\approx 1$. The picked-up particle initially follows a cycloid in the $xz$ plane (starting along $z$ and then getting deflected along $x$), with a relativistic drift in the $x$ direction $u\sim\sqrt{\sigma}$ and a comparable 4-velocity of gyration in the drift frame. 
} 
as well as particles accelerated at X-points. The particle loses its gyration energy on the IC cooling timescale, and its oscillating $z$-momentum eventually vanishes. Then the cold particle simply follows the field-line drift along $x$, with the bulk speed of the unstructured outflow.

%%%%%%%%%%%%%%%%%%%%%%%%%%%%
\begin{figure}
\centering
\resizebox{1.05\hsize}{!}{\includegraphics{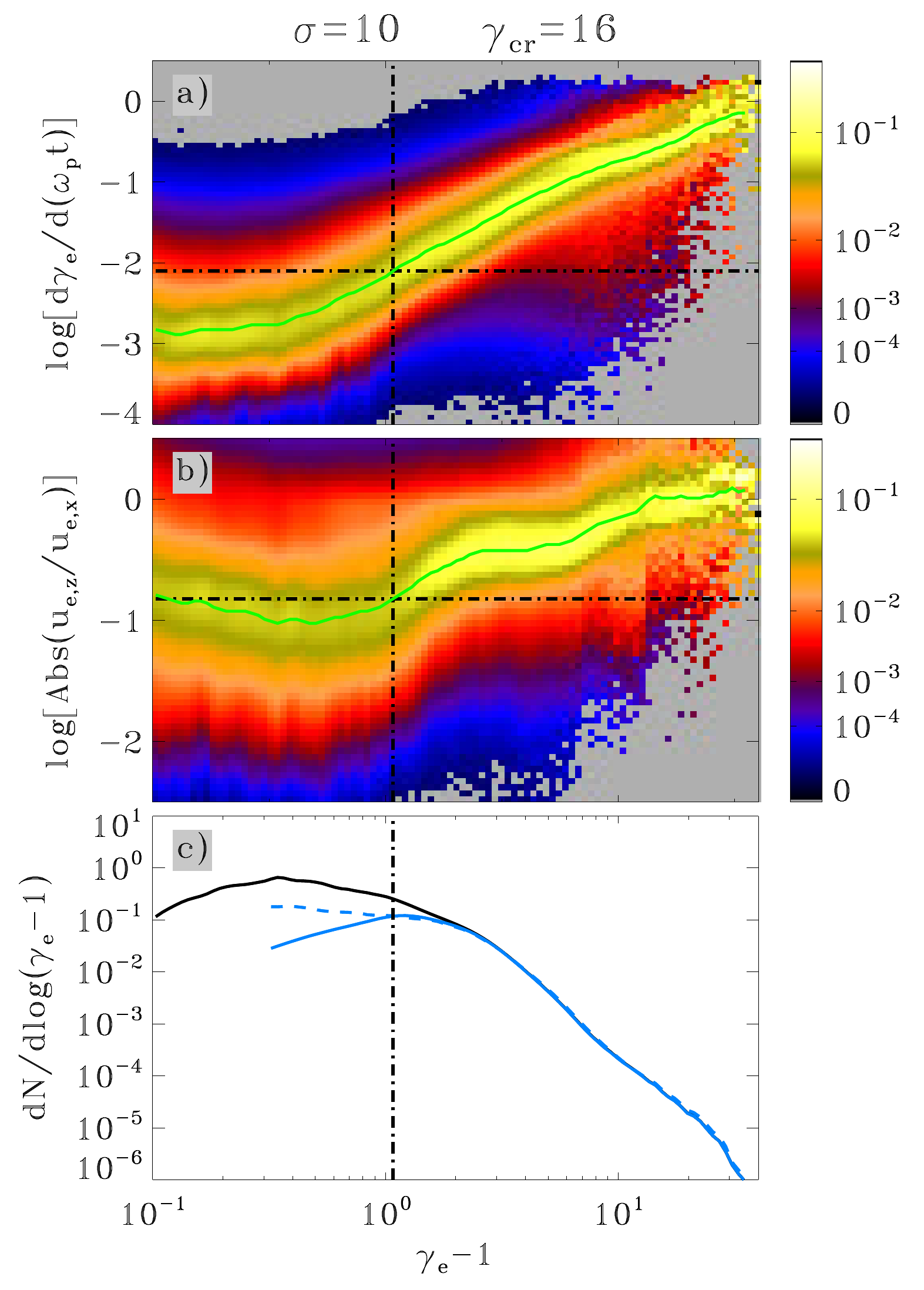}}
 \caption{Diagnostics of  
 particle acceleration in the radiative simulation ($\gammacr=16$).
 The top and middle panels show 2D histograms of the tracked particles, time-averaged in the interval $1.5\lesssim ct/L\lesssim 5$. 
In the top panel, the vertical axis 
represents
the median 
acceleration rate $d\gamma_e/d(\omega_{\rm p} t)$ during the 
recent time interval $\Delta t$ equal to  
the IC cooling time of the particle $t_{\rm IC}(\gamma_e)$. 
In the middle panel, the vertical axis 
represents
the median value of $|u_{e,z}/u_{e,x}|$ 
during the recent $\Delta t =t_{\rm IC}(\gamma_e)$.
The histogram values are shown by color. They are normalized separately for each bin $\delta\ln(\game-1)$ (the integral over each vertical slice equals unity); the green curve shows the median of the histogram.
The vertical dashed line is the boundary ($\game>2$) of one of the two high-energy intervals discussed in the text.
Bottom panel: contributions of particles with strong recent accelerations (solid blue curve) or high $z$-momenta (dashed blue curve) to the particle energy distribution in the reconnection layer (black curve). 
Strong acceleration/high $z$-momentum is defined as the region above the horizontal dashed line in the top/middle panel.
One can see that both conditions define particles that dominate the distribution at $\game>2$.
  }
 \label{fig:spectime}
\end{figure}
%%%%%%%%%%%%%%%%%%%%%%%%%%%%

The pick-up process is quick --- it occurs on the gyration timescale 
$\sim\gamma/(\eta_{\rm rec}\sqrt{\sigma}\, \omega_{\rm p})$,  comparable to the acceleration timescale at X-points. The sharp rise in energy, with the development of a significant $z$-momentum, clearly identifies the acceleration episodes when analyzing the histories of particles.
For a particle with a Lorentz factor $\game$ at a given moment of time, we examine its recent history during one cooling timescale $t_{\rm IC}(\game)$, and measure the evolution of $d\game/dt$ during the past $\Delta t=t_{\rm IC}$. Its median value represents a characteristic gain rate $d\game/dt$ in the recent past. It is expected to be high for particles with high $\game$, as otherwise they would have cooled.
In addition, we measure the ratio $|u_{e,z}/u_{e,x}|$ for the particle during the same time interval $\Delta t=t_{\rm IC}$, and compute its median value. 
Particles that have just experienced the pick-up or X-point acceleration are expected to have a high $|u_{e,z}/u_{e,x}|$, 
as 
their gyration involves a large $z$-component.
In order to accumulate good statistics, we repeat the measurement of $d\game/dt$ and $|u_{e,z}/u_{e,x}|$ for a large number of particles (roughly 10 million), with a temporal cadence of $18\,\omega_{\rm p}^{-1}$. 

The results are presented in \fig{spectime}. 
In the $\game-d\game/dt$ plot we observe a transition at $\game\sim 2$, from particles with a low recent acceleration $d\game/d(\omega_{\rm p} t)\sim 10^{-3}$ to particles with a strong recent acceleration. A similar transition is seen in the plot of $|u_{e,z}/u_{e,x}|$. Particles with $\game<2$ have a dominant $u_{e,x}$, as expected for cold particles moving with the local bulk velocity of the unstructured outflow. The ratio $|u_{e,z}/u_{e,x}|$ quickly increases at $\game>2$ to values comparable to unity. In particular, $|u_{e,z}/u_{e,x}|\sim 0.5$ for $3\lesssim \game\lesssim 7$ and $|u_{e,z}/u_{e,x}|\sim 1$ for $\game\gtrsim 7$.
These diagnostics provide further confidence in the separation of the 
energy
distribution in the reconnection region into two distinct components: bulk motion-dominated ($\game\lesssim 2$) and high-energy particle injection ($\game\gtrsim 2$).

%############################################
\begin{figure}
\centering
\resizebox{1.05\hsize}{!}{\includegraphics{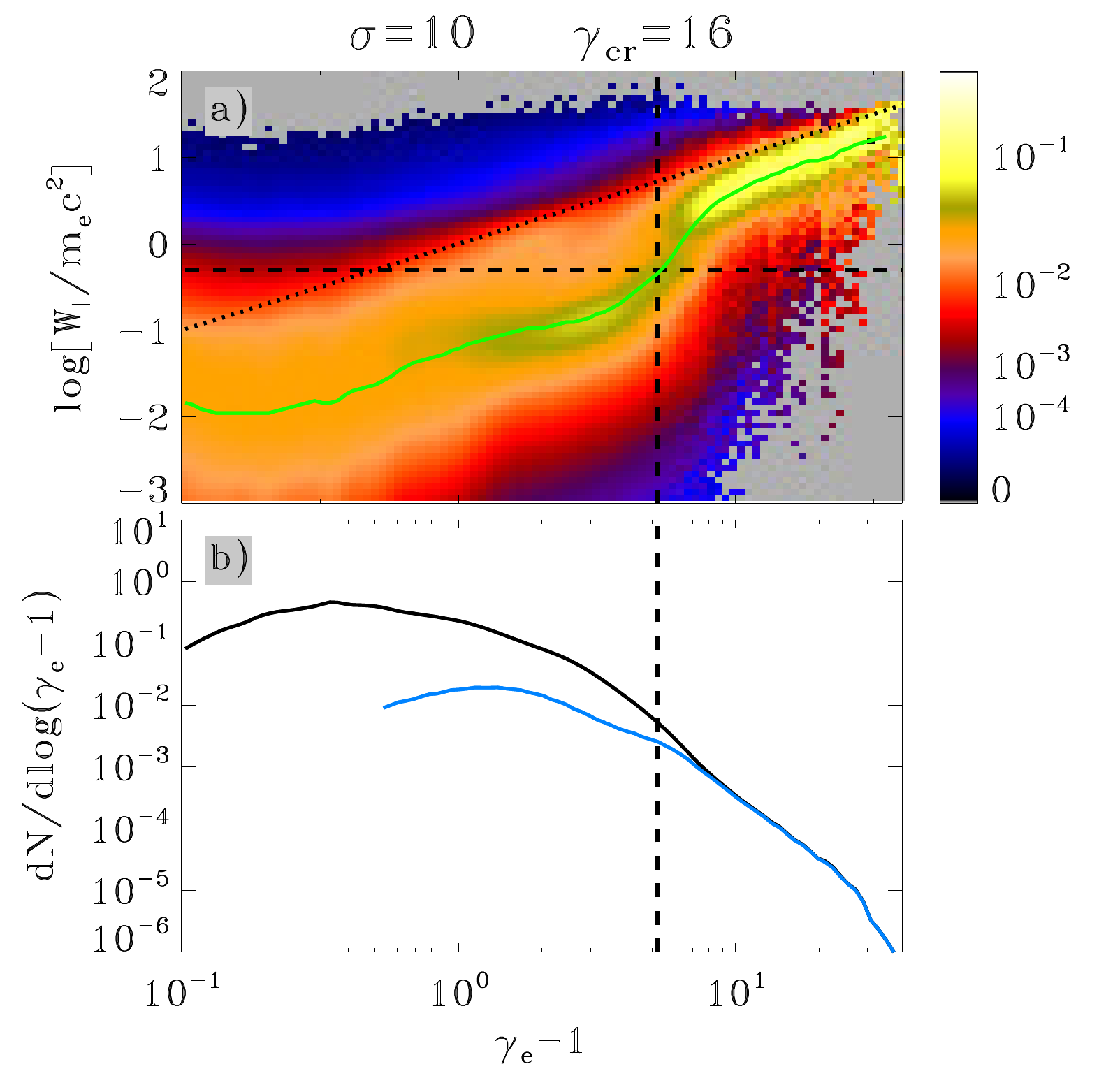}}
 \caption{Diagnostics of 
 particle acceleration by the non-ideal electric field $E_\parallel$ in the radiative simulation ($\gammacr=16$).
 In the top panel, the vertical axis represents  
the cumulative work $W_\parallel$ done on the particle by 
 $E_\parallel$ during the recent time interval $\Delta t$ equal to the IC cooling time $t_{\rm IC}(\gamma_e)$. The color-coded 2D histogram was constructed and normalized similarly to that in \fig{spectime}, with the green curve showing the median value. For particles with $\game\gg 6$, $W_\parallel$ accounts for a large fraction of their energy, approaching the relation $W_\parallel=(\gamma_e-1)m_e c^2$ shown by the dotted black line.
Bottom panel: contribution of particles with large $W_\parallel$ (blue curve) to the particle energy distribution in the reconnection layer (black curve). 
Here large $W_\parallel$ is defined as the region above the horizontal dashed line in the top panel.
One can see that the non-ideal acceleration dominates the distribution at $\game>6$.
}
 \label{fig:specwork}
\end{figure}
%############################################

Further insight into the acceleration mechanism at $\game\gtrsim 6$ is provided by \fig{specwork}. For a particle with Lorentz factor $\game(t)$, we calculated the ``parallel'' work done 
during the recent time interval $\Delta t=t_{\rm IC}(\game)$,
\be\label{eq:wpar}
W_\parallel(t)=e\int_{t-t_{\rm IC}[\gamma_e(t)]}^t E_\parallel\, v_{e}^{\parallel} \,dt,
\ee
where $E_\parallel =\bmath{E}\cdot \bmath{B}/B$ and $v_{e}^{\parallel} =\bmath{v}_e\cdot \bmath{B}/B$ are the electric field and the particle velocity components parallel to the local magnetic field $\bmath{B}$.
The condition of ideal magnetohydrodynamics $E_\parallel=0$ is maintained 
almost everywhere in the reconnection layer except the X-points, where a large $E_\parallel$ develops and non-ideal effects become important. Thus, the work $W_\parallel(t)$ is a proxy for recent particle acceleration at X-points. 
As one can see from \fig{specwork}, $W_\parallel$ is large for particles with $\game>6$ and becomes the main factor responsible for their high energies.
We conclude that the high-energy end of the particle spectrum 
is the result of 
acceleration
by $E_\parallel$ at X-points. The energy released through the X-point acceleration approximately equals the energy received (and radiated) by the population with $\game>6$; it accounts for $\sim 1\%$ of the total energy released by magnetic reconnection.

%############################################
\begin{figure}
\centering
\resizebox{1\hsize}{!}{\includegraphics{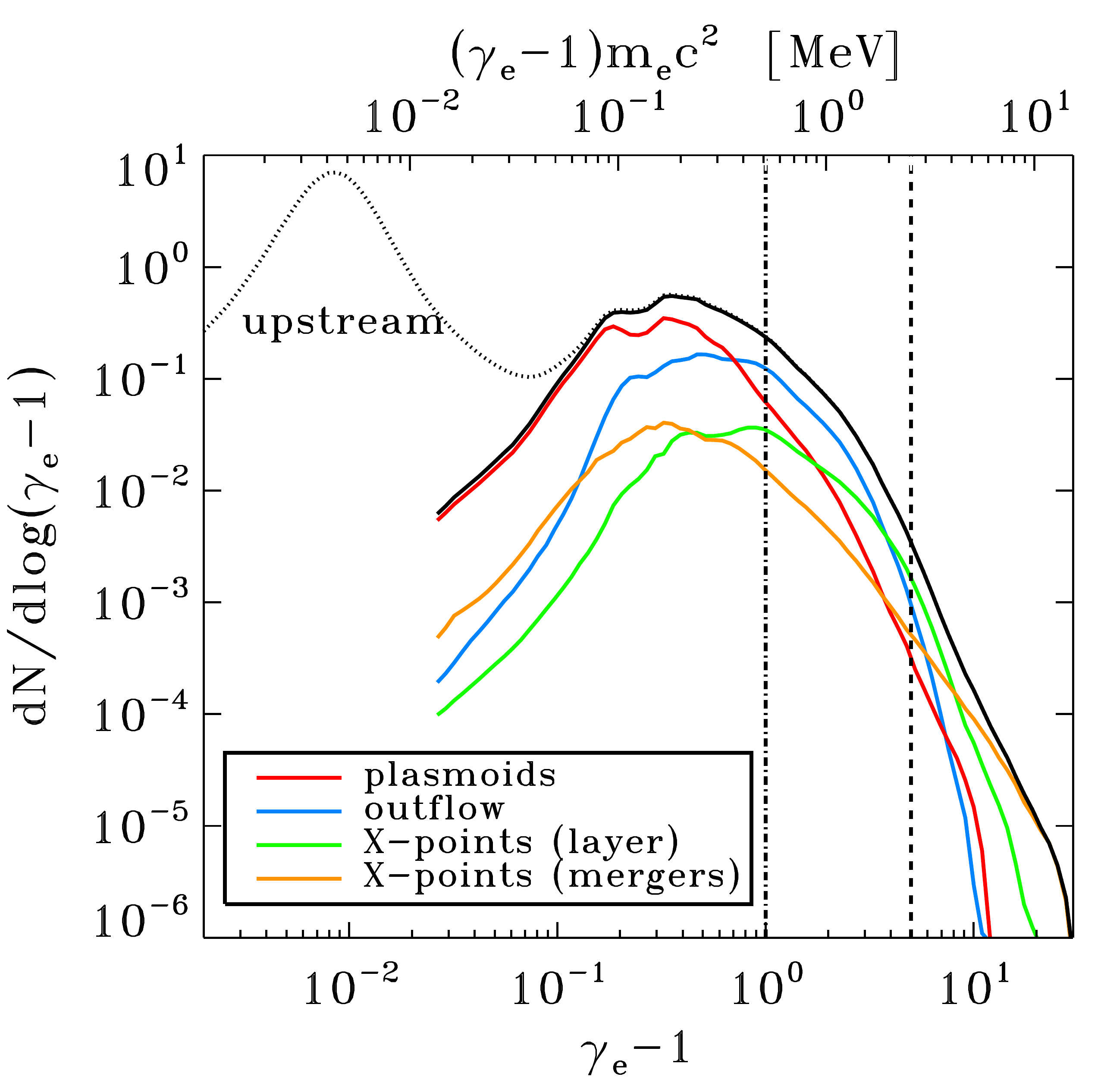}}
 \caption{
 Energy distribution of particles in the radiative simulation
 ($\gammacr=16$), time-averaged over the interval $1.5\lesssim ct/L\lesssim 5$.
 The black solid curve shows the distribution in the reconnection region (same as in \fig{spectime} and \fig{specwork}), 
 and the dotted  black curve shows the distribution in the entire box, including the inflow region. 
The distribution in the reconnection region 
is further dissected into four components. This dissection is performed using particle spectra binned in vertical spatial slices of the reconnection region. Each slice has the thickness of 100 grid cells (i.e., $20\comp$) along the $x$ direction,  and it extends in the $y$ direction until the boundaries of the reconnection region (as identified by the mixing criterion described in \sect{plasm}). We also use the magnetic vector potential to identify the locations of X-points and the contours of plasmoids (see \citealt{sironi_16}). If an X-point lies at the boundary between two neighboring plasmoids, it is identified as an X-point formed by 
plasmoid mergers, and the particle spectrum from that spatial slice contributes to the yellow curve. Otherwise, the X-point resides in the main reconnection layer, and the corresponding spectrum contributes to the green curve. Spatial slices that contain only particles residing inside plasmoids give the red curve. If none of these conditions is met, the slice intersects the unstructured outflow located in between plasmoids (and not containing X-points), which gives the blue curve.
 }
 \label{fig:specslice}
\end{figure}
%############################################

A more detailed analysis demonstrates two types of X-point acceleration: most particles with $6\lesssim \game\lesssim 10$ are 
generated
in the reconnection plane $y\approx 0$, and particles with $\game>10$ are mainly generated in the secondary (vertical) reconnection layers formed at the interfaces between merging plasmoids. 
The higher energies achieved by particles in the secondary layers 
reflect the stronger magnetization: 
the magnetic field in plasmoids is stronger than in the inflow region, and so a merger-induced reconnection layer effectively has a higher 
$\sigma$ than the nominal $\sigma=10$.

In summary, we find that three distinct populations contribute to the spectrum in the reconnection region: ({\it i}) at
moderate  
energies, $\game<2$, the particle distribution is dominated by bulk motions of large  plasmoids cooled to non-relativistic temperatures 
and pushed by magnetic stresses to mildly relativistic speeds, 
({\it ii}) at intermediate energies, $2\lesssim \game\lesssim 6$, the distribution is dominated by particles freshly picked up by the unstructured outflows from X-points toward neighboring plasmoids, and ({\it iii}) at the highest energies, $\game\gtrsim 6$, the distribution becomes dominated by particles accelerated by $E_\parallel$ at X-points, either in the main reconnection layer or in the reconnection layers formed at the interface between merging plasmoids. The contributions of all these acceleration mechanisms are shown in \fig{specslice}. The populations ({\it ii}) and ({\it iii})  account for $\fHE\sim 20\%$ of the total power dissipated by reconnection.

\begin{figure}
\centering
\resizebox{\hsize}{!}{\includegraphics{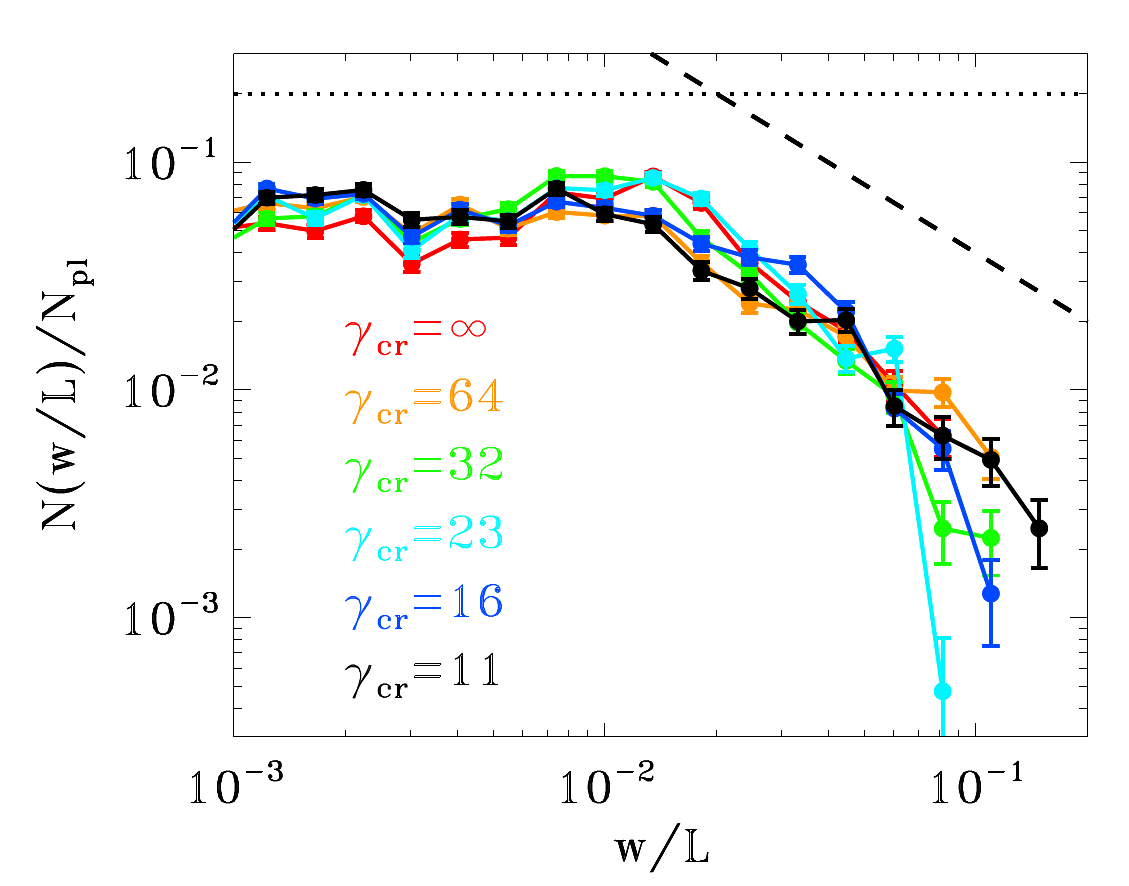}}
 \caption{Cumulative distribution 
 of plasmoid width $w$, for different values of $\gammacr$ as indicated in  the 
 figure.
 The histogram (with Poissonian error bars) is normalized to the overall number of plasmoids $N_{\rm pl}$. The corresponding differential distribution 
 is given by
 $f(w)=dN(w)/dw$. The predictions $N(w)\propto w^{-1}$ by \citet{uzdensky_10,loureiro_12} and $N(w)\propto {\rm const}$ by \citet{huang_12} are plotted as dashed and dotted black lines, respectively.}
 \label{fig:islstat}
\end{figure}

\subsection{Dependence on the radiation field intensity}
\label{sec:scan}

\begin{figure}
\centering
\resizebox{\hsize}{!}{\includegraphics{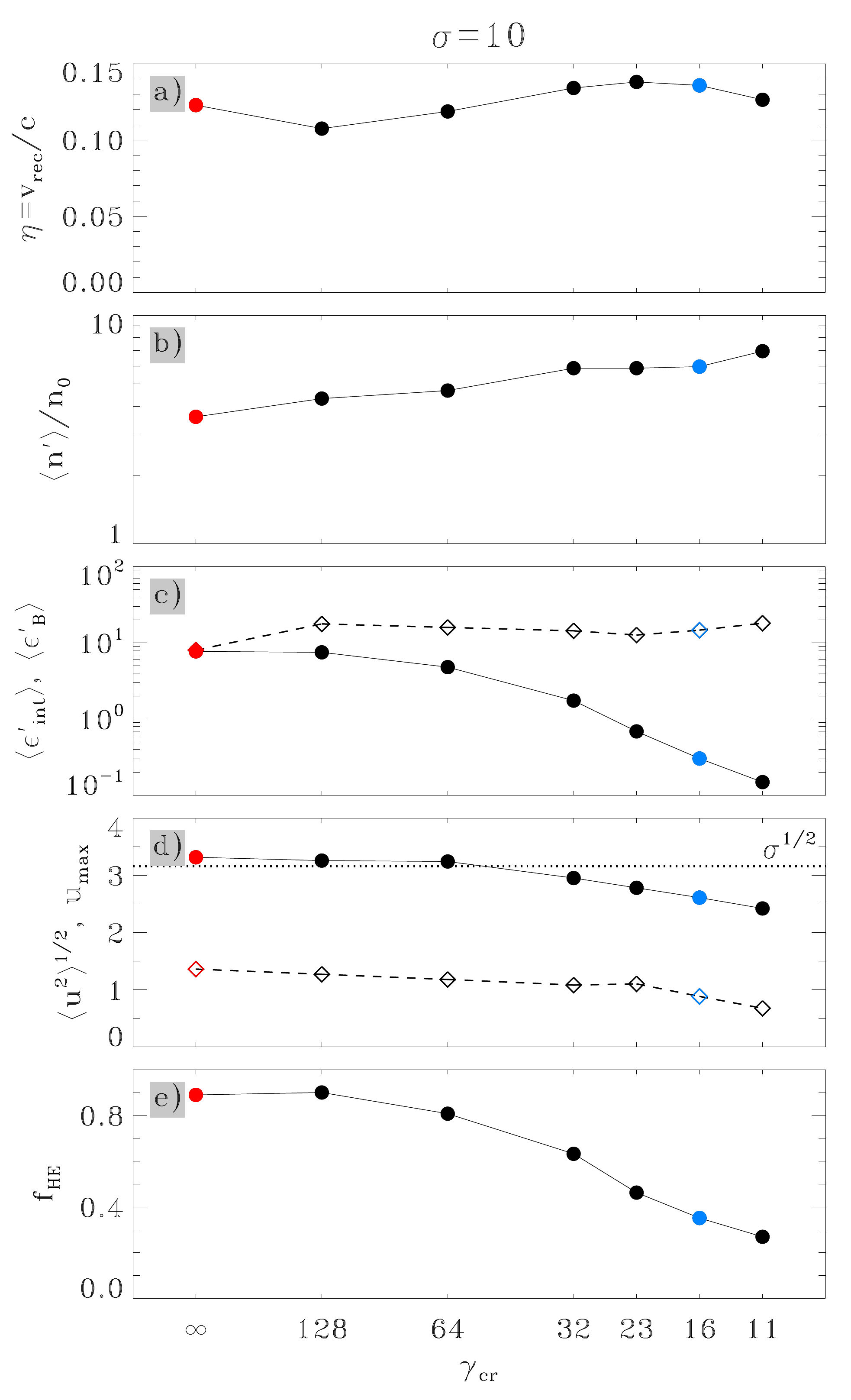}}
 \caption{Dependence of the main properties of the reconnection system on radiative losses. The decreasing parameter $\gammacr$ from left to right corresponds to increasing the level of radiative losses. Red and blue colors highlight the two simulations discussed in most detail in the text: $\gammacr=\infty$ (non-radiative) and $\gammacr=16$ (strongly radiative).  (a) The reconnection rate, or inflow rate, in units of the speed of light; (b) the  average comoving particle density in the reconnection region (as defined in \sect{plasm}), in units of the upstream density $n_0$; (c) the comoving magnetic energy fraction $\epsilon'_{B}$ (dashed) and internal energy fraction $\epsilon'_{\rm int}$ (solid), averaged over the reconnection region (they are both normalized to the upstream rest mass energy density); (d) the root mean squared bulk momentum $\langle u^2\rangle^{1/2}$ in the reconnection region (dashed) and the maximum outflow bulk momentum $u_{\rm max}$ (solid), with  the horizontal dotted black line corresponding to the Alfv\'enic limit of $\sqrt{\sigma}$; (e) the ratio of the power emitted by the high-energy particles (i.e., the component that cannot be accounted for by bulk motions; in \fig{specfit}, it is shown as the green dashed line ) and the net power emitted by all particles in the reconnection region.
 For the comoving quantities in panels (b) and (c), for each computational cell we transform particle and field quantities to the corrresponding fluid rest frame, given the local bulk velocity $\bmath{\beta}$. All points are obtained by time averaging over $1.5\lesssim ct/L\lesssim 5$.}
 \label{fig:allfluids}
\end{figure}

In the previous sections we compared the results of two simulations:
the fiducial model with strong cooling ($\gammacr=16$) and the non-radiative model ($\gammacr=\infty$). We have also run simulations with different values of $\gammacr$ to study in more detail the effect of increasing radiative losses (decreasing $\gammacr$  from $\infty$ down to 11.3). Th results are presented in \fig{islstat} and \fig{allfluids}.

\fig{islstat} shows  the cumulative distribution of plasmoid sizes. It demonstrates that
the statistics of plasmoid sizes is not appreciably affected by the level of cooling losses.

\fig{allfluids} shows that the reconnection rate is nearly independent of the degree of IC losses. Furthermore, the plasma compression in the reconnection layer is only moderately increased by cooling: from $\langle n^\prime\rangle/n_0\sim 4$ in the non-radiative case to $\langle n^\prime\rangle/n_0\sim 7$ in the case of $\gammacr=11.3$. This is in agreement with the results of \citet{werner_19}, but in contradiction with earlier works \citep{uzdensky_11,uz_14,begue_16}, which argued that cooling results in strong compression until the plasma pressure in the reconnection layer balances the upstream magnetic pressure. 
Instead, we find that the dynamic, magnetically dominated plasmoid chain sustains the required pressure with a negligible plasma contribution, close to the force-free regime, and thus avoids strong compressions.
Most of the magnetic energy in the layer is contained in the plasmoids, and the resulting mean magnetic energy in the reconnection layer is nearly insensitive to IC losses.

%%%%%%%%%%%%%%%%%%%%%%%%%%%%
%%%%%%%%%%%%%%%%%%%%%%%%%%%%

By contrast, the internal energy in the reconnection region drops by two orders of magnitude as $\gammacr$ decreases from infinity to $\gammacr=11.3$. The internal energy fraction $\epsilon'_{\rm int}$ is defined as the average internal energy density in the reconnection region normalized to the rest-mass energy density of the upstream plasma. The internal energy density was evaluated in each computational cell by boosting the plasma energy density from the lab frame to the plasma rest frame (which has the local velocity $\bmath{\beta}$), using the approximation of isotropic plasma stress tensor in the rest frame.\footnote{See \citet{sironi_16} for further details. In a few cases, we have checked that a Lorentz transformation of the full stress energy tensor to the plasma rest frame yields nearly identical results (i.e., that our values are not sensitive to the assumption of isotropic stress tensor).}

%%%%%%%%%%%%%%%%%%%%%%%%%%%%
\begin{figure*}
\centering
\resizebox{0.75\hsize}{!}{\includegraphics{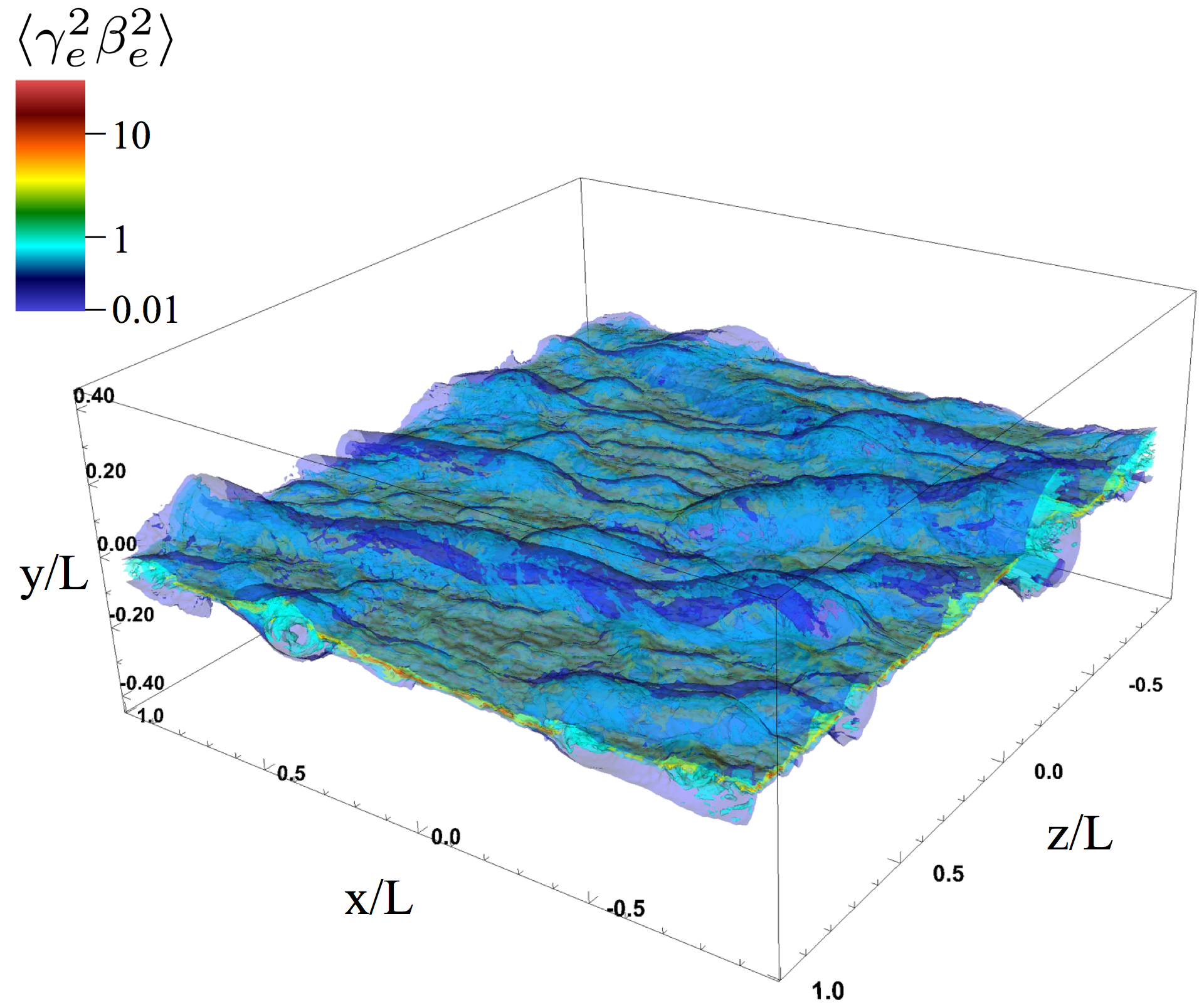}}
 \caption{\rev{3D structure of the reconnection layer at time $t=4.6\,(L/c)$ in our largest 3D run ($L_z=L=806\comp$) with very strong IC losses  ($\gamma_{\rm cr}=11.3$). 
 We show the region $|y|/L<0.45$ where reconnection occurs 
(the actual extent of the computational box along $y$ grows with time as described in \sect{setup}). We show the local average $\langle\gamma_e^2\beta_e^2\rangle=\langle\gamma_e^2\rangle-1$, obtained for each cell by averaging over the particles in the neighboring $5\times5\times5$ cells.
This quantity is proportional to the local {\it total} IC power radiated per particle.  The figure shows four isosurfaces of $\langle\gamma_e^2\beta_e^2\rangle$, each colored according to the value of  $\langle\gamma_e^2\beta_e^2\rangle$.
One can see that large plasmoids are cold (blue and cyan), whereas the thin regions of the reconnection layer, where the freshly energized plasma has not cooled yet, are sites of bright IC emission (green and orange). 
An animation of this figure is available. The video shows the time evolution from $t=0$ to $t=5.6\,L/c$. The animation is also available at \url{http://user.astro.columbia.edu/~lsironi/Site/share/ic_in_reconnection/}.}
}
 \label{fig:fluid3d}
\end{figure*}
%%%%%%%%%%%%%%%%%%%%%%%%%%%%

The bulk motions in the reconnection layer somewhat slow down at low $\gammacr$, because Compton drag opposes the accelerating magnetic forces. However, this effect is much weaker than the loss of internal energy, in agreement with analytical estimates in \sect{plasm}.
In agreement with expectations, $\gamma\beta$ stays below the $\sqrt{\sigma}$ limit, i.e. plasma motions in the layer are moderately sub-Alfv\'enic, even for the smallest and fastest plasmoids.

\fig{allfluids} also shows the measured $\fHE$, the ratio of the power emitted by the high-energy particles (i.e., the component that cannot be accounted for by bulk motions) and the net power emitted by all particles in the reconnection region. 
It drops from $\fHE \sim 90$\% at $\gammacr=\infty$ to $\fHE\lesssim 30$\% at low $\gammacr$, when strong radiative losses limit the lifetime of accelerated particles.

%%%%%%%%%%%%%%%%%%%%%%%%%%%%
\begin{figure}
\centering
\resizebox{\hsize}{!}{\includegraphics{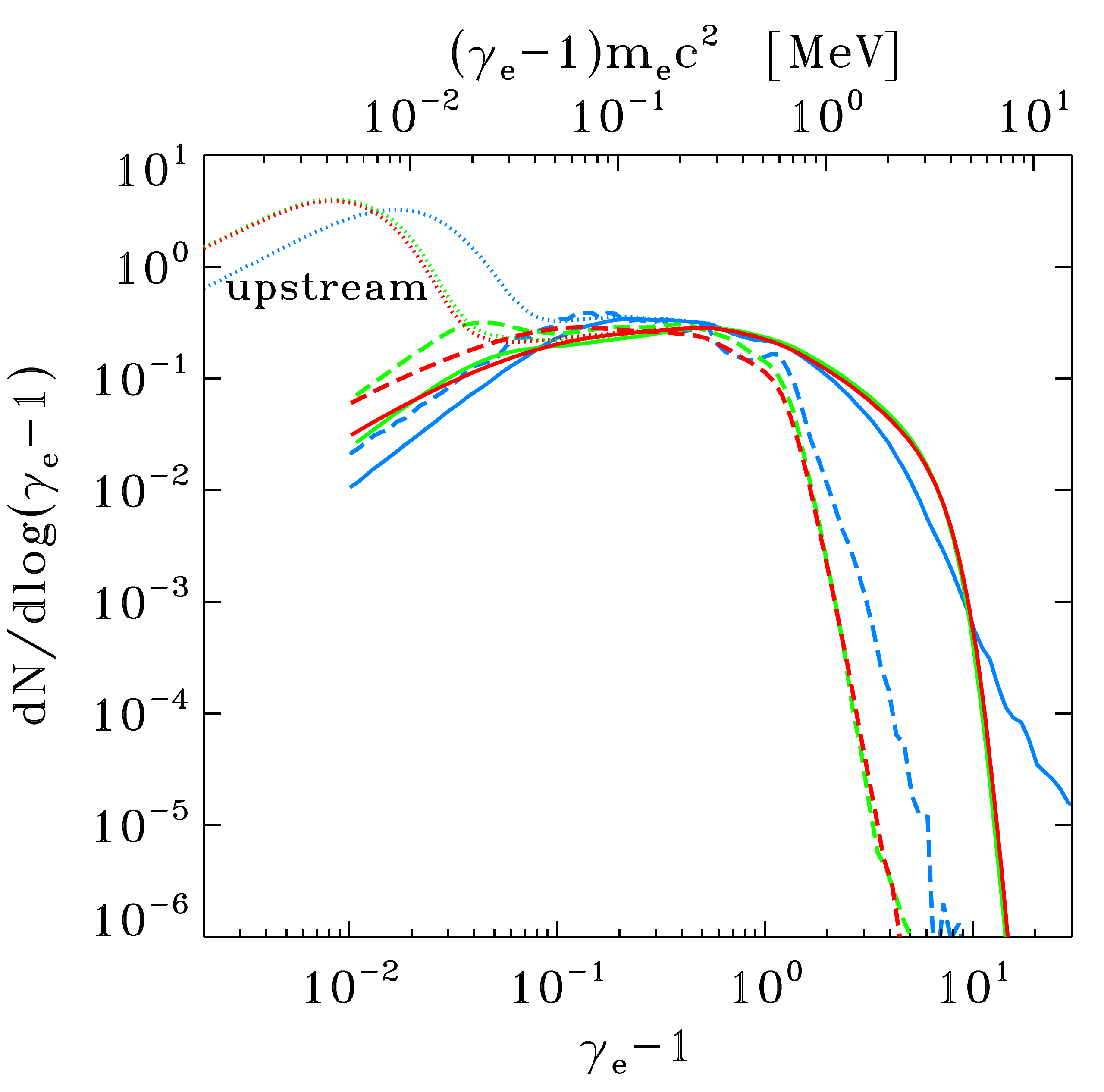}}
 \caption{\rev{{Particle  
 \AB{distribution $dN/d\log(\gamma_e-1)$ in the 3D and 2D simulations}
with $\gammacr=11.3$, time-averaged in the interval $1.5\lesssim ct/L\lesssim 5.8$.
\AB{For each simulation, the distribution was calculated in the reconnection region (solid curves) and in the entire computational box (dotted curves). The dashed curves}
show the distribution in the reconnection region when accounting only for {\it bulk} kinetic energy (i.e., we plot $dN/d\log(\gammab-1)$ as a function of $\gammab-1$).  The color coding represents different box half-lengths along the $z$ direction: $L_z=0$ in cyan (i.e., a 2D simulation),  
\AB{and}
$L_z=115\comp$ in green,   $L_z=806\comp$ in red. The box half-length along the $x$ direction is 
$L=806\comp$
\AB{in all the models.} 
The distributions are normalized to the number of particles in the reconnection region.
}}}
 \label{fig:spec3d}
\end{figure}
%%%%%%%%%%%%%%%%%%%%%%%%%%%%

%%%%%%%%%%%%%%%%
%\vspace{0.3in}
\section{
3D Simulations
}
\label{sec:3d}

\rev{
\AB{Real plasmas are not two-dimensional, i.e. the invariance in the $z$-direction is not enforced. Then the kink instability develops in the current sheet \citep{zenitani_07,ss_14}. When the 3D reconnection process is viewed in 
the $x$-$y$ cross section, it looks similar to that in 2D simulations --- plasmoids are formed by the tearing instability and accelerated by magnetic stresses along the $x$-axis. The 2D plasmoids are the cross sections of magnetic flux ropes that extend in the $z$-direction. The kink instability bends the flux ropes, forcing them to entangle and reconnect.}

\AB{The characteristic coherence length of a dynamic flux rope in the $z$ direction is comparable to its thickness (the plasmoid size in the $x$-$y$ plane), and relativistic particles traverse this length on the light crossing timescale. 
Thus, unlike the 2D simulations, the particles trapped in a flux rope are not completely buried. They get a chance to become repeatedly energized as they move along the bent flux rope, approach new reconnection sites, and interact with the fast outflows emanating from X-points.
The repeated energization events may better counteract radiative losses. 
}

\AB{To explore the effects of 3D dynamics on radiative reconnection we have performed 3D simulations with a setup similar to that of our 2D simulations.
Due to computational constraints, we choose a computational box with half-lengths in the $z$ and $x$ directions of $L_z=L_x=L=806\comp$.}
We employ periodic boundary conditions in the $z$ direction. 
The skin depth is resolved with 2.5 cells, and we initialize the upstream plasma with one particle per computational cell. We keep $\sigma=10$ and $B_g/B_0=0.1$, \AB{same} as in our fiducial 2D runs. In order to (marginally) preserve the hierarchy of relevant scales and energies discussed in \sect{params}, we \AB{focus on the model with strongest cooling, which has $\gammacr=11.3\gtrsim\sigma$; then} 
the relation in \eq{lbox} still marginally holds.

\AB{
The results are presented in \fig{fluid3d} and \fig{spec3d}. 
Similar to our 2D runs} 
(cf. panel (c) of \fig{fluid2}),
we find that most of the reconnection region (in particular the large plasmoids\AB{/flux ropes}) is
\AB{efficiently cooled down (\fig{fluid3d}).}
\AB{
The hottest plasma (and hence strongest IC emission) is}
localized in small regions where individual particles are quickly energized at  X-points and in the fast outflows emanating from the X-points. This is clearly observed in the thin 
\AB{parts}
of the reconnection layer, where the freshly energized plasma has not cooled yet. The same happens at the interfaces between merging 
\AB{flux ropes}
where secondary current sheets are formed.

In \fig{spec3d} we compare the particle 
\AB{energy distributions found in the 3D and 2D simulations with the same magnetization and cooling rate. One can see that two 3D simulations with different box sizes $L_z$ gave nearly identical results, indicating convergence, and these results are similar to those of the 2D simulation.}
\AB{Particle acceleration is somewhat more efficient in the 3D simulation in the range $3\lesssim \gamma_e\lesssim 10$; 
however, this effect is modest.}\footnote{\rev{At higher energies, $\gamma_e\gtrsim 10$, the difference between 2D and 3D results may be attributed to the difference in merger dynamics of 2D plasmoids versus 3D flux ropes. However, particles with $\gamma_e\gtrsim 10$ make a  negligible contribution to the overall particle census and the  IC power.}}
Regardless of the dimensionality 
of the computational domain, 2D or 3D,
the 
peak of the 
\AB{energy}
distribution ($\gamma_{e}\lesssim 2$) is controlled by 
\AB{the}
bulk motions, i.e. the particles populating the peak are cold. 
%The bulk energy spectra are nearly the same between 2D and 3D. 
}

%%%%%%%%%%%%%%%%
%\vspace{0.3in}
\section{
Conclusions
}
\label{sec:disc}

Kinetic plasma simulations provide an efficient tool to study the nonlinear dynamics of magnetic reconnection. This paper focused on the radiative regime, where heated particles radiate their energies on a timescale much shorter than the light-crossing time of the reconnection layer. \revv{Our simulation setup was motivated by the hard X-ray activity of accreting black holes, which can be powered by radiative reconnection in magnetically dominated regions of the corona.} \revvv{In this paper, we studied radiative reconnection in a pair-dominated plasma with a high magnetization parameter $\sigma\gg 1$, which is likely found near the black hole spin axis.}

For simplicity, we fixed the radiation field during the simulation and assumed that it is composed of low-energy photons that scatter in the Thomson regime (see also \citealt{werner_19}). 
As the main parameter controlling radiative (Compton) cooling we chose $\gammacr$ (defined in \sect{cd}), which is proportional to the cooling timescale and inversely proportional to the radiation density. We have studied reconnection with various levels of radiative losses by perfoming simulations with different $\gammacr$, from $\gammacr=\infty$  (non-radiative) down to $\gammacr=11.3$ (extremely radiative). Running the simulations in two dimensions allowed us to use large computational boxes, with length $2L$ up to $13440 \comp$. This permits good separation of spatial and energy scales, which is particularly important in the radiative regime. 
\rev{\AB{We have also performed a 3D simulation for the case of $\gammacr=11.3$. Its results suggest that our main conclusions from the 2D simulations will hold in 3D models.} }
\\

Our 
% main 
conclusions are as follows:
\bi
\item 
In radiative magnetic reconnection, plasma particles are energized as they enter the reconnection layer, then quickly cool and become locked in  plasmoids at a low temperature $kT\ll 100$~keV. \revvv{This makes thermal Comptonization unable to generate 100 keV X-rays.}
\item
The plasma energy in the reconnection layer (and its inverse Compton emission) is dominated by mildly relativistic {\it bulk} motions of cold plasmoids, which are pulled against Compton drag by magnetic stresses. This confirms the radiative mechanism described in B17. We find that, \revv{for a magnetically dominated pair plasma}, the plasmoid bulk motions have the effective temperature $\sim 100$~keV and their inverse Compton emission mimics thermal Comptonization.
\item
Particle acceleration in radiative reconnection is less efficient compared with the non-radiative regime. However, there are accelerated particles in the reconnection layer, which form a high-energy tail of the particle distribution function. We have identified two mechanisms injecting energetic particles: acceleration at X-points and particle pick-up by outflows from X-points. Both mechanisms are nearly impulsive (i.e. operate on a timescale much shorter than the cooling timescale). Slower acceleration processes, in particular Fermi reflection and magnetic compression \citep[e.g.,][]{guo_19,petropoulou_18} are suppressed in the radiative regime. We found that about 20\% of the dissipated reconnection power goes into high-energy particles. Their inverse Compton emission will generate a high-energy tail of the radiation spectrum, which may explain the MeV tail detected in the hard state of Cyg X-1  \citep{mcconnell_02}.
\item 
Radiative losses influence the picture of plasma dynamics in the reconnection layer: radiation efficiently removes thermal plasma pressure, exerts Compton drag on bulk motions, and clears cavities inside moving plasmoids (\fig{fluid2}). However, these dynamic effects weakly change the net reconnection rate and magnetic field structure in the reconnection layer. Our results support the conjecture of B17 that the magnetic forces moving plasmoids are approximately the same in the radiative and non-radiative regimes. Furthermore, the statistics of plasmoid sizes are similar (\fig{islstat}). We conclude that the plasmoid chain formation is largely controlled by magnetic stresses, and plasma pressure plays a minor role even in the hot (non-radiative) regime. We also note that similar plasmoid chains were observed in ``force-free'' simulations that completely neglected plasma pressure and inertia \citep{parfrey_15}, again illustrating the dominant role of magnetic stresses in shaping the structure of relativistic reconnection. 
\ei

In summary, our first-principle simulations support radiative magnetic reconnection \revv{in the magnetically dominated regime} as the mechanism powering the hard state of accreting black holes. Our results confirm the radiative mechanism proposed in B17, give the effective temperature of bulk motions in the reconnection layer $kT_{\rm b}\sim 100$~keV, and give the fraction of dissipated reconnection power deposited into high-energy particles, $\fHE\sim 20$\% (much lower than in previous non-radiative simulations). 

\revv{In all our models reconnection occurred in $e^\pm$ plasma with magnetization $\sigma=10$. Recent studies of non-radiative reconnection in electron-proton and electron-positron-proton plasmas have shown that for $\sigma\gtrsim1$, leptons still pick up a significant fraction of the dissipated magnetic energy \citep{rowan_17,rowan_19,werner_18,petropoulou_19}. Future work will explore how the outcome of radiative reconnection depends on $\sigma$ and ion abundance.}

It may also be useful to investigate how the results change in the presence of a strong guide field, stronger than $0.1B_0$ assumed in our simulations.  
We also leave for future work detailed calculations of the X-ray spectrum. Such calculations can be done similarly to the Monte-Carlo simulations in B17 but with the particle energy distribution provided by first-principle kinetic plasma simulations.
\\

LS acknowledges support from DoE DE-SC0016542, NSF ACI-1657507, NASA ATP NNX17AG21G and NSF PHY-1903412. A.M.B. is supported by NASA grant NNX17AK37G, Simons Foundation grant \#446228, and the Humboldt Foundation. The simulations have been performed at Columbia (Habanero and Terremoto), and with NASA-HEC (Pleiades) and NERSC (Cori) resources.

\bibliography{blob}
\end{document}